\documentclass{article}

\usepackage{arxiv}

\usepackage[utf8]{inputenc} % allow utf-8 input
\usepackage[T1]{fontenc}    % use 8-bit T1 fonts
\usepackage{hyperref}       % hyperlinks
\usepackage{url}            % simple URL typesetting
\usepackage{booktabs}       % professional-quality tables
\usepackage{amsfonts}       % blackboard math symbols
\usepackage{nicefrac}       % compact symbols for 1/2, etc.
\usepackage{microtype}      % microtypography
\usepackage{lipsum}
\usepackage{graphicx}
\usepackage{enumitem}
\usepackage{gensymb}

\usepackage{float}

\usepackage{mathtools}

\makeatletter
\newcommand*{\textlabel}[2]{%
  \edef\@currentlabel{#1}% Set target label
  \phantomsection% Correct hyper reference link
  #1\label{#2}% Print and store label
}
\makeatother

\usepackage[ruled,linesnumbered]{algorithm2e}
\makeatletter
\newcommand{\nosemic}{\renewcommand{\@endalgocfline}{\relax}}% Drop semi-colon ;
\newcommand{\dosemic}{\renewcommand{\@endalgocfline}{\algocf@endline}}% Reinstate semi-colon ;
\let\oldnl\nl% Store \nl in \oldnl
\newcommand{\nonl}{\renewcommand{\nl}{\let\nl\oldnl}}% Remove line number for one line
\makeatother

\usepackage[style=numeric-comp, sorting=none]{biblatex}
\title{Optical imaging of MHD bubble flow \\ in Hele-Shaw liquid metal cells}

\author{
  Aleksandrs Jegorovs\\
  Institute of Numerical Modelling\\
  University of Latvia (UL)\\
  Riga, Latvia, Jelgavas 3, 1004 \\
  \texttt{aleksandrs.jegorovs@lu.lv} \\
  \And
  Mihails Birjukovs\\
  Institute of Numerical Modelling\\
  University of Latvia (UL)\\
  Riga, Latvia, Jelgavas 3, 1004 \\
  \texttt{mihails.birjukovs@lu.lv} \\
  \And
  Jevgenijs Telicko\\
  Institute of Numerical Modelling\\
  University of Latvia (UL)\\
  Riga, Latvia, Jelgavas 3, 1004
  \And
  Andris Jakovics\\
  Institute of Numerical Modelling\\
  University of Latvia (UL)\\
  Riga, Latvia, Jelgavas 3, 1004
  }

\begin{document}
\maketitle

\begin{abstract}

As a simple and affordable alternative to often prohibitively expensive or unavailable X-ray and neutron imaging, an improved optical imaging method for bubble flow in Hele-Shaw liquid metal cells is presented, enabling measurements with a significantly greater liquid metal layer thickness than previously reported. This enables studying bubble dynamics with varying degrees of geometric confinement, without or with applied magnetic field. The main principles and the experiment setup, as well as the necessary image/data processing pipeline are described, and preliminary results show that the proposed methods can be used to quantify the effects of varying gas flow rate and magnetic field configuration on bubble chain flow in a Hele-Shaw cell.

\end{abstract}


\section{Introduction}
\label{sec:intro}

Multiphase flow in external magnetic field (MF) and its control is a topic of great importance for industrial applications and in fundamental science research. Two-phase metal/gas flows are especially relevant in the field of metal processing. Prime examples of bubble flow in liquid metal are metal stirring and impurity removal, continuous metal casting \cite{casting-euler-les, limmcast, embr-visualized, embr-cift, casting-new-collective-dynamics-models, casting-lagrange-bubbles, particles-plus-bubbles-simulations, particles-multiscale-simulations, particle-aggregation-simulations, udv-bubbles-visualization, taborda-les-euler-lagrange,stirring-cfd-srm, metal-strring-2, metal-strring-3, metal-strring-4, simulations-experiments-dual-jets, particles-plus-bubbles-simulations, particles-multiscale-simulations, particle-aggregation-simulations}. These and similar processes are so common that any form of their improvement is always desired. Bubble flow parameters, such as bubble number density, their velocity and surface shapes are important factors, which determine how bubbles interact with MF, metal flow, each other, and inclusions/impurities. To improve industrial systems and enable their reliable control using applied MF, there should be a firm grasp of fundamental physics of the underlying processes, and good empirical models should be available.

Initial understanding of bubble flow physics stems from analysis of a single free bubble (unconfined, no MF). Hydrodynamic mechanisms (e.g., path instabilities) of a raising bubble are currently rather well-understood, as these were actively studied using numerical modelling, as well as experiments, mostly in transparent liquids, which enable use of convenient optical methods \cite{prl-path-instability,natcomms-shape-dynamics,spiral-to-zigzag-explained,spiral-to-zigzag-explained-2,shape-and-wake-simulations,uttt-path-instability,clift-bubbles, in-depth-study-of-ellipsoid-kinematics, optics-collective-dynamics, 2021-review-article-bubbles-in-liquid-metal, gaudlitz-shape-wake-variations-1-bubble, optics-collective-dynamics}. Of course, such simplification is generally inapplicable for liquid metals due to their opacity, and one cannot study the effects of MF in water, since its electrical conductivity is several orders of magnitude below that of relevant metals. Flows of single MHD free bubble within external MF were also studied in detail experimentally and through numerical models \cite{zhang-mf-simulations,zhang-mf-vertical,zhang-thesis,hzdr-ibm-bubbles-thesis,hzdr-bubbles-mf, uttt-x-ray-single-bubble, uttt-path-instability, udv-wake-flow-structure,udv-longitudinal-field, udv-review-article, udv-transverse-field, udv-bubbles-visualization, x-ray-prime-code, dns-longitudinal-field, zhang-mf-vertical, imb-transverse-field}, though not as often -- these studies show how transversal and longitudinal ML configuration affect bubble path instability and bubble's shape.

Bubble interactions occur via various mechanisms. As a base case, two bubble systems are studied -- the main mechanisms of interaction are momentum exchange via induced vortices and interaction with the fluid pressure field. Depending on the relative bubble position, different interaction scenarios can take place. In case of vertically aligned ascending bubbles, the trailing bubble is generally attracted to the leading one, behind which pressure drop occurs. Meanwhile, vortices generated by the leading bubble may act to destabilize the path of the trailing bubble and potentially result in its ejection from the pursuit trajectory \cite{prl-path-instability,shape-and-wake-simulations, hzdr-ibm-bubbles-thesis, x-ray-prime-code, kusuno-bubble-pair-coupling-2021, bubbles-side-by-side, bubbles-in-line-2020, hele-shaw-bubble-vortex-interaction}. Bubble coalescence and breakup depends on pressure inside bubbles, their relative velocity and even duration of contact, as well as the surrounding flow field. In magnetohydrodynamic (MHD) bubble flow, coalescence and breakup events can be expected to be strongly affected by MF, since inter-bubble film drainage and surrounding flow field should be damped by the Lorentz force \cite{hele-shaw-coalescence, tomiyama-og-gangster-hele-shaw-bubble-coalescence, hzdr-bubbles-mf}.

Both single and two bubble cases, however, are still not representative of industrial processes where many bubbles ascend and interact together in jets and columns. It therefore makes sense to transition to the next level of approximation and physical analysis, and consider bubble chains, where bubbles are injected via a nozzle one after another with a fixed frequency. Bubble chains entail the relevant collective dynamics and instabilities, but are still simple enough to experiment with and model numerically, including how the application of MF affects them \cite{hzdr-ibm-bubbles-thesis, baakeNeutronRadiographyVisualization2017, x-ray-bubble-chain-simulate, x-ray-prime-code, x-ray-bubble-breakup, x-ray-bubble-coalescence, x-ray-validation, megumi-x-rays, birjukovsArgonBubbleFlow2020, birjukovsPhaseBoundaryDynamics2020, birjukovs2021resolving}.

Major advances towards studying MHD bubble flow were made for continuous casting, where systematic research was conducted \cite{casting-new-collective-dynamics-models, casting-euler-les, casting-euler-musig, casting-lagrange-bubbles, embr-cift, taborda-les-euler-lagrange}, although beyond that the available research for MHD bubble flow is relatively lacking. While numerical modeling is a prominent segment of multiphase MHD flow research, it should be noted that capturing bubble surface explicitly in high detail is very computationally challenging, if not prohibitive in some instances. It is possible to avoid the need of resolving bubbles fully by utilizing reduced bubble flow models. A popular approach is tracking bubbles as point-like particles via Lagrangian models \cite{casting-lagrange-bubbles, casting-euler-les, casting-euler-musig, taborda-les-euler-lagrange, casting-new-collective-dynamics-models}. This method allows modelling high number density bubble flow with relative ease. However, since  inter-bubble interactions are not resolved explicitly, they are treated with empirical models, which, while already good enough, still could use significant improvement, especially where MF effects on collective dynamics are concerned.

The arsenal of experimental methods that can be applied in industrial settings is very limited. System size, high temperature and chemical aggressiveness of metals are the main constraints. Therefore, it is common to instead use reduced scale models for systems of interest and conduct research in an easier to control environment, given that dimensionless numbers for the downscaled systems are sufficiently well-matched to those of the original system. Temperature and chemical hazards are mitigated by using metals or metal alloys that are liquid at near room temperature, such as gallium (Ga) \cite{baakeNeutronRadiographyVisualization2017, birjukovsArgonBubbleFlow2020, birjukovsPhaseBoundaryDynamics2020, birjukovs2021resolving} and gallium-indium-tin (GaInSn) \cite{strumpfExperimentalStudyRise2017, uttt-path-instability, udv-bubbles-visualization, udv-review-article}. However, metal opacity is still a problem. Unlike experiments with purely hydrodynamic bubble flow, where pure water can often be used, and optical transmission methods are feasible, there is no known substitute for liquid metal that is both transparent and has a similar order of magnitude of electrical conductivity. Therefore, MHD bubble flow must be studied in a liquid metal medium, and the issue of opacity must be solved by other means.

Study of flows in opaque liquids is possible and selection of methods quite extensive. The main goal is to retrieve bubble positions and shapes, with the best possible spatial and temporal resolution. Compatibility with MF application is also required. Therefore, non-invasive contactless methods are of interest. Acoustic methods, such as use of an array of ultrasonic Doppler Velocimetry (UDV)\cite{strumpfExperimentalStudyRise2017,udv-bubbles-visualization,udv-longitudinal-field,udv-review-article,udv-transverse-field,udv-wake-flow-structure}, can produce one- or two-dimensional flow velocity field, from which bubble positions can, in principle, be inferred, but only very roughly. UDV does not have the necessary spatial resolution. The ultrasound transit time technique (UTTT) is also an acoustic method which allows detecting bubble positions, but also has low spatial resolution and does not allow recovering bubble shapes, and shape irregularity can also introduce errors to position estimates \cite{uttt-path-instability,uttt-x-ray-single-bubble}. Contactless inductive flow tomography (CIFT) is a promising method for 3D velocity field reconstruction, but it is not yet clear how to apply it to non-disperse bubble flow, and it also does not allow for precise bubble localization \cite{cift-1, cift-2, embr-cift}. Therefore, UDV and CIFT are relevant mostly as complementary techniques for other methods that could resolve bubble shapes.

Much more viable experimental methods for study of bubble MHD flow in opaque metal are radiographic transmission contrast-based methods. The principle is similar to optical measurements of transparent media, but X-rays and neutrons are used instead. Due to different attenuation/scattering coefficients for metal and gas, the transmitted intensity map captured after passing though the metal reveals bubble locations in 2D projection images. Currently, these methods are X-Ray radiography (XR), ultra-fast X-ray computed tomography (UXCT), and neutron radiography (NR). XR and NR are similar, with the main difference being that X-rays interact with atom electrons versus with nuclei in the case of neutrons. Both methods have been demonstrated as viable for studying bubble flow without and with applied MF \cite{baakeNeutronRadiographyVisualization2017,birjukovsArgonBubbleFlow2020,birjukovsPhaseBoundaryDynamics2020,birjukovs2021resolving,x-ray-bubble-breakup,x-ray-bubble-coalescence,x-ray-validation, saito-neutrons-1, saito-neutrons-2,x-ray-bubble-chain-simulate, megumi-x-rays}. The produced 2D images do not enable the 3D shape recovery, even though bubble thickness in the beam direction can be deduced from images. Depending on the X-ray/neutron beam flux, liquid metal thickness and scintillation screens, images show different values of signal-to-noise and contrast-to-noise ratios (SNR and CNR, respectively), associated with degraded image quality, which is especially evident when thick metal layers are imaged \cite{birjukovsArgonBubbleFlow2020, birjukovsPhaseBoundaryDynamics2020, birjukovs2021resolving}. There exists a maximum thickness (on the order of tens of millimeters) beyond which measurement analysis becomes impossible for a given experimental setup. This is because one must have a large enough field of view (FOV) and frame rate to capture bubble trajectories and shape dynamics. UXCT is a method used to rapidly scan flow system cross-sections and has been successfully used to reconstruct 3D bubble shapes from accumulated slices at high frame rates ($\mathcal{O}(10^3)~\text{Hz}$). Although UXCT has seen quite a few application instances \cite{x-ray-ct-1,x-ray-ct-2,x-ray-ct-3}, it is yet unclear how, if possible, it can be made compatible with applied MF, given that X-rays are produced by an electron beam. It is also worth noting that recent developments in ultrasound computed tomography (UCT) are very promising, and UCT has already been applied to bubble flow measurements in liquid metal, and is readily compatible with MF application \cite{sven-ultrasonic-tomography, sven-ultrasonic-tomography-2}. While UCT has a smaller temporal resolution, it currently offers a better reported spatial resolution than UXCT.

Experiments with thin (Hele-Shaw) liquid metal flow systems (i.e., thin vessels) are promising for studying bubble collective dynamics in particular \cite{maxworthy-og-gansta-hele-shaw, tomiyama-og-gangster-hele-shaw-bubble-coalescence, hele-shaw-bubble-swarm-1, hele-shaw-bubble-swarm-2, hele-shaw-bubble-vortex-interaction, hele-shaw-coalescence, hele-shaw-high-re-bubbles}, since thicker vessels make it hard to force bubble interaction even at higher flow rates. Geometric confinement readily promotes bubble breakup, coalescence and, importantly, agglomeration (i.e., motion in closely packed groups), and one can assess the effects of MF on the above processes/events \cite{x-ray-bubble-breakup,x-ray-bubble-coalescence}. However, such Hele-Shaw systems are still opaque, and while XR and NR are easier to apply for thinner metal layers, some of the issues persist. The problem with NR is that neutron flux activates the metal sample. NR is incompatible with GaInSn, since one of its components, indium, has a very long half-life, and long continuous measurements require a considerable supply of metal -- neither Ga nor GaInSn are particularly cheap, which also applies to some of the other relevant low melting point metals/alloys. In addition, NR requires specialized equipment which is tremendously expensive, and is only available at a few facilities like the Paul Scherrer Institute \cite{psi-facility,kaestner2011} and the Oak Ridge National Laboratory \cite{oak-ridge-neutrons}. Beam time at such neutron sources can also be challenging to acquire, in general, due to extremely high demand and overbooking. While XR does not have the material activation issue and does not require the infrastructure scale associated with NR, the X-ray tubes along with a shielded laboratory and other necessary hardware are still an investment that many institutions and research groups cannot afford.

Despite not being feasible for thicker systems, optical imaging has been attempted for bubble flow in Hele-Shaw liquid metal cells before -- however, the thickness was limited to $\sim 1 ~mm$, and single bubble flow was studied \cite{KlaasenHeleShawMetalSolo}. The main issue is that above a certain thickness, optical transmission of bubbles does not occur, since even thin metal films are still opaque. In this paper, we offer an optical imaging method that sidesteps this issue, enabling bubble flow imaging for systems with $\sim 3\text{-}4~mm$ liquid metal thickness. The proposed setup is readily compatible with applied MF, which we demonstrate by optically imaging bubble flow at different gas flow rates and with various MF configurations, and showing that the effects of varying MF and gas flow rate can be quantified based on the acquired images. The experimental and data processing methodology developed here should also be applicable to similar experiments with other opaque liquids.

\section{Experiments}

\begin{figure}[htbp]
\begin{center}
\includegraphics[width=0.925\textwidth]{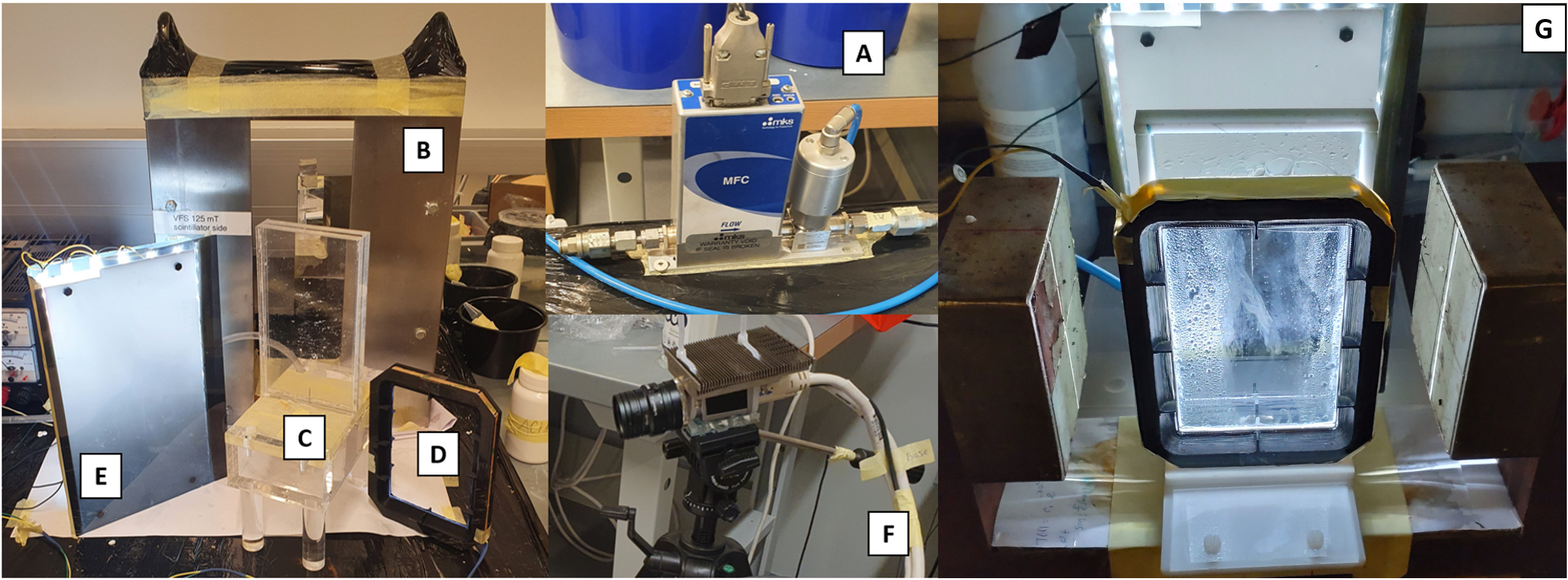}
\caption[]{Experimental setup: (A) gas flow rate controller, (B) MF system, (C) glass/acrylic vessel, (D) side illumination, (E) backlighting, (F) camera. (G) shows the setup in its assembled state.}
\label{fig:exp}
\end{center}
\end{figure}

Normally, optical imaging in liquid metal, or another opaque liquid, relies on raising bubbles maintaining contact with both cell walls (normal to the backlighting) simultaneously, i.e. bubbles can be viewed as tunnels through the liquid metal layer (optical projections) \cite{KlaasenHeleShawMetalSolo, hele-shaw-optics-magnetic-fluid}. The cell is, of course, optically transparent. Bubble transparency can be achieved by either by maintaining a high bubble volume by controlling gas flow rate, or by reducing Hele-Shaw cell thickness. Since the goal is to minimize the thickness limitation and measure at a wide range of desirable flow rates, these are not viable options. In addition, bubble dimension across the cell gap should be large enough for buoyancy force to overcome drag due to wall contact. Our Hele-Shaw cell combines mineral glass (glass, for brevity) for the front- and backplates and with extruded acrylic glass (glass, for brevity) for the rest of the structure, to accommodate for gas inlet ports. Glass and acrylic are both transparent and chemically inert (for our purposes). Glass is more scratch resistant, while acrylic is much easier to machine, and both are readily available and cheap. The vessel's internal dimensions are $160 \times 88 \times 3~ {mm}^3$ (Figure \ref{fig:exp}C) and are mainly limited in height and width to fit inside a permanent magnet system (Figure \ref{fig:exp}B), but at the same time allow capturing sufficiently long bubble trajectories. It is important that metal does not wet the vessel walls, because even the thinnest metal films between cell walls and bubbles are optically opaque.

Since there is no material restriction due to neutron radiation, argon and GaInSn (melting point well below room temperature) are used, and thus a heating system is not required, as would be the case with pure Ga. However, GaInSn is slightly more expensive than Ga. Argon because of its inertness, accessibility and affordability. Argon is introduced via a hypodermic needle (diagonally cut cross-section) inserted vertically into the bottom plate of the vessel and sealed with a rubber O-ring, compressed by a plastic screw. The gas flow rate is regulated using an \textit{MKS Instruments} mass flow controller (MFC, calibrated for argon, shown in Figure \ref{fig:exp}A) capable of maintaining flow rates within the range of $50$-$2000~sccm$ (standard cubic centimeters per minute).

Ga in GaInSn, however, still reacts with oxygen in ambient air, forming oxides which are deposited at the free surface on top of the metal cell, and then also on bubble surfaces and vessel walls, as the oxides are mixed into the cell volume. Oxidized surfaces have substantially different properties compared to pure alloy, i.e. surface tension, contact angle and glass/acrylic wetting. Oxide deposition on the front-/backplates of the vessel is especially challenging for optical measurements. Deposition happens continuously, from grayish transparent overlay up to completely opaque film, accumulating in a matter of minutes. To prevent oxide formation, we use a hydrochloric acid (HCl) solution ($26~\%$). Perhaps counterintuitively, despite intensifying GaInSn aeration and oxide/metal mixing, increased gas flow rate actually helps to remove wall bound oxides, preventing film formation. More details regarding issues and troubleshooting related to oxidation can be found in \ref{appendix:A}.

\bigskip

\begin{figure}[htbp]
\begin{center}
\includegraphics[width=0.325\textwidth]{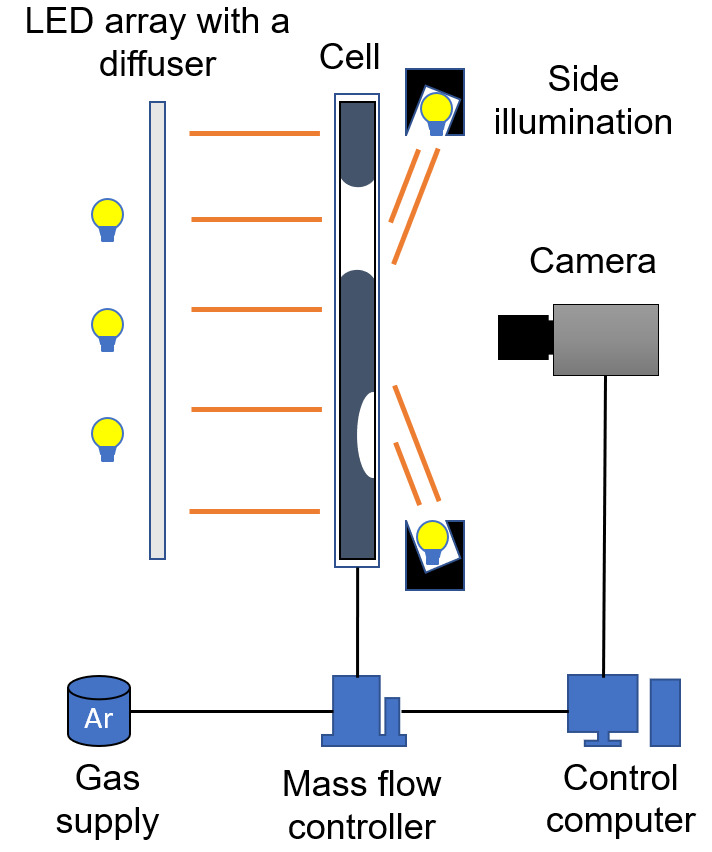}
\caption{A schematic representation of the experimental setup shown in Figure \ref{fig:exp}.}
\label{fig:scheme}
\end{center}
\end{figure}

\begin{figure}[htbp]
\begin{center}
\includegraphics[width=0.8\linewidth]{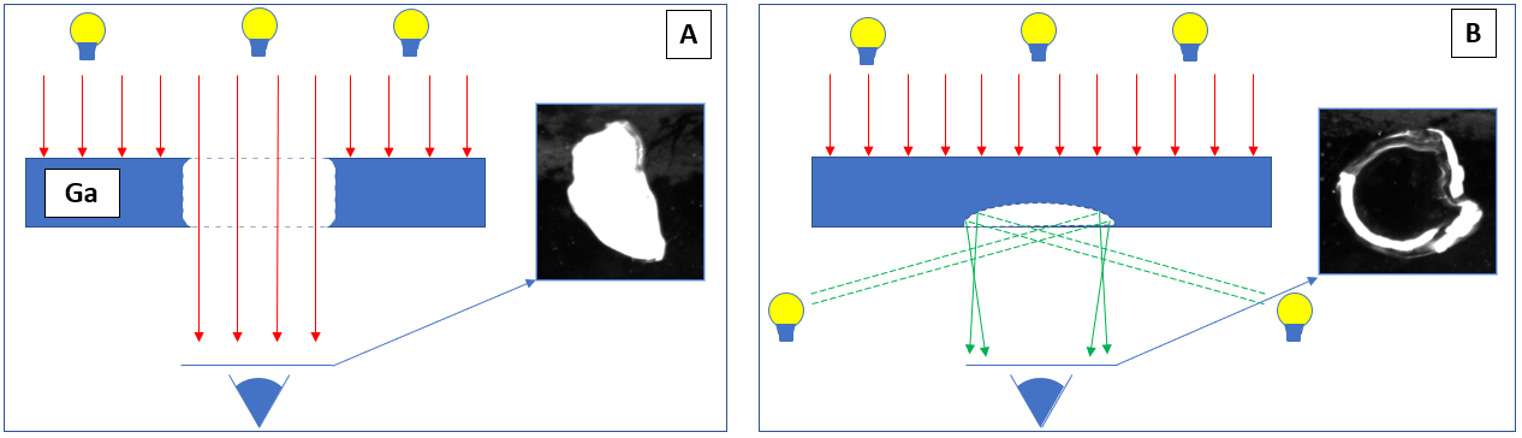}
\caption{(A) A translucent bubble illuminated by the backlight, and (B) a contour of a bubble in contact with a single wall highlighted by the side illumination.}
\label{fig:exp_illum1}
\end{center}
\end{figure}

A homogeneous backlight (Figure \ref{fig:exp}E) consisting of a simple grid of LED diodes and a diffuser is placed behind the vessel to create a homogeneous illumination, so that fully translucent bubbles are clearly visible. However, this is where one runs into the main issue -- some bubbles ascend exclusively along the front-/backplate and cannot be captured via transmitted light (Figure \ref{fig:scheme}). To address this, side illumination (Figure \ref{fig:exp}D) can be used, directed from along the metal vessel perimeter at angles such that it only illuminates the edges of the bubble-wall interface, which will ideally manifest as rings in the FOV, as shown in Figure \ref{fig:exp_illum1}. Performing recordings in a darkened room reduces the number and impact of stray light reflections from the vessel walls and mirror-like liquid metal surface inside it. The possible issues are that, first, closer to the FOV center, the bubble edges may only be partially illuminated, and second, small acid droplets may form and get stuck at the edges of the metal vessel, possibly contributing bright reflections.

\clearpage

An important constraint on the side illumination system design is the small thickness of the metal vessel, which means that one must have a high number density of the light sources to illuminate the bubble edges. Here, we used a white LED strip with $\sim 10^2$ LEDs per meter, spaced equidistantly. Selective illumination from the desired angles is achieved using a specifically shaped 3D-printed reflector frame. The geometry files are available on \textit{GitHub} at \href{https://github.com/ajegorovs/optical_mhd_hele_shaw_cell}{ajegorovs/optical\_mhd\_hele\_shaw\_cell}, and are the two parts that are locked together with the LED strip inside.

For image acquisition, we used a \textit{Basler acA2000-340km} camera (grayscale, $2040 \times 1088~ {px}^2$, CMOS, 350 frames per second, FPS -- Figure \ref{fig:exp}F) with a PCI-E \textit{Euresys Grablink Full} capture card. The camera is compatible with a variety of native optics, and we chose the \textit{Kowa} LM16HC F1.4 f16mm 1" lens to perform recordings at a $\sim30~ cm$ distance from camera to the vessel center (thickness) and to have FOV with as much of the image space as possible filled with the vessel. A short focus distance introduces a significant fish-eye effect, which can be removed after calibration imaging with reference checkerboard grids. It must be noted that coupling the \textit{Basler} camera with the \textit{Euresys} capture card was not possible out-of-the-box, because the built-in software does not enable image sequence recording. To perform recordings, the \textit{Euresys} camera operation script for single frame capture was modified to include RAM reservation for temporary sequence storage and iterative capture. The code for image acquisition is available on \textit{GitHub} at \href{https://github.com/ajegorovs/Grablink-Full-sequence-acquisition/tree/master}{ajegorovs/Grablink-Full-sequence-acquisition}. Frame timing (camera FPS) was tested by filming a diode array, with its diodes activated in a sequence using an \textit{Arduino} board. FPS was verified by analyzing the diode activation duration and activation order. However, for other camera/card choices, the above may not be an issue. More details are provided in \ref{appendix:B}.

The setup as described above is readily reproducible, since it is simple, cheap, and the necessary materials are widely available. This makes the proposed experiment feasible for research groups and/or individual researchers with smaller budgets.

\section{Image processing}

The entire data processing pipeline is implemented in \textit{Python} environment, since it is a free scripting language, and is sufficiently fast for the purposes of this paper. The main libraries used for this project are \href{https://opencv.org/}{\textit{OpenCV}} (image processing), \href{https://networkx.org/documentation/stable/index.html#}{\textit{NetworkX}} (mathematical graphs), \href{https://docs.python.org/3/library/itertools.html}{\textit{itertools}} (combinatorics) and \href{https://scipy.org/}{\textit{SciPy}} (data interpolation). The developed code is open-source and is available on \textit{GitHub} at \href{https://github.com/ajegorovs/bubble_process_py}{ajegorovs/bubble\_process\_py}.

\subsection{Bubble detection}

\begin{figure}[htbp]
\begin{center}
\includegraphics[width=0.85\linewidth]{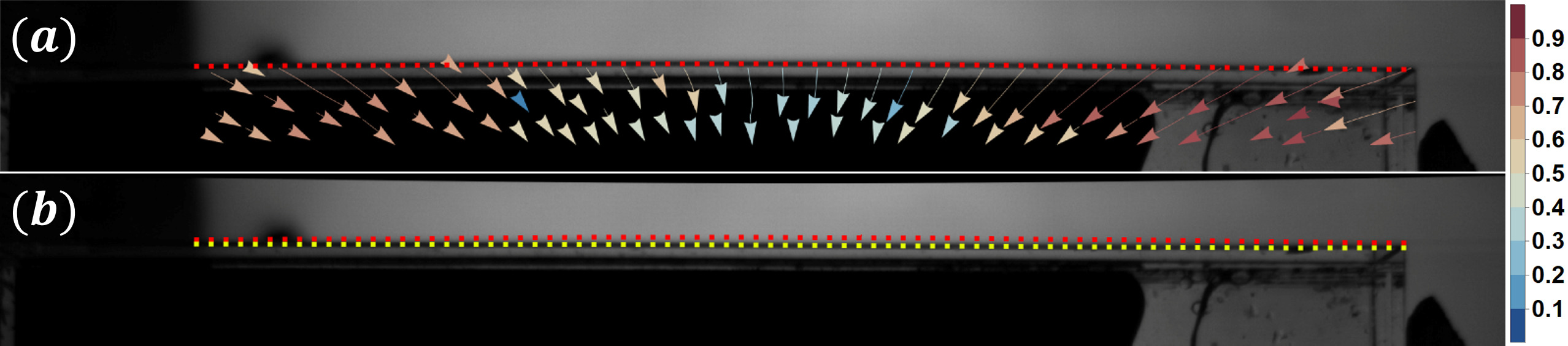}
\caption{Fish-eye optical distortion correction: segments of (a) raw and (b) corrected images. Red contours indicate the original shape/position of the vessel boundary, the yellow contour shows the vessel boundary after image correction. The correction displacement field inside the vessel is shown with exaggerated streamlines, color coded by the relative magnitude. Displacement field estimated via image keypoint tracking using combined SURF \cite{surf-ipol, surf-og} and BRISK \cite{brisk-og} methods.}
\label{fig:calibration}
\end{center}
\end{figure}

\begin{figure}[htbp]
\begin{center}
\includegraphics[width=0.80\linewidth]{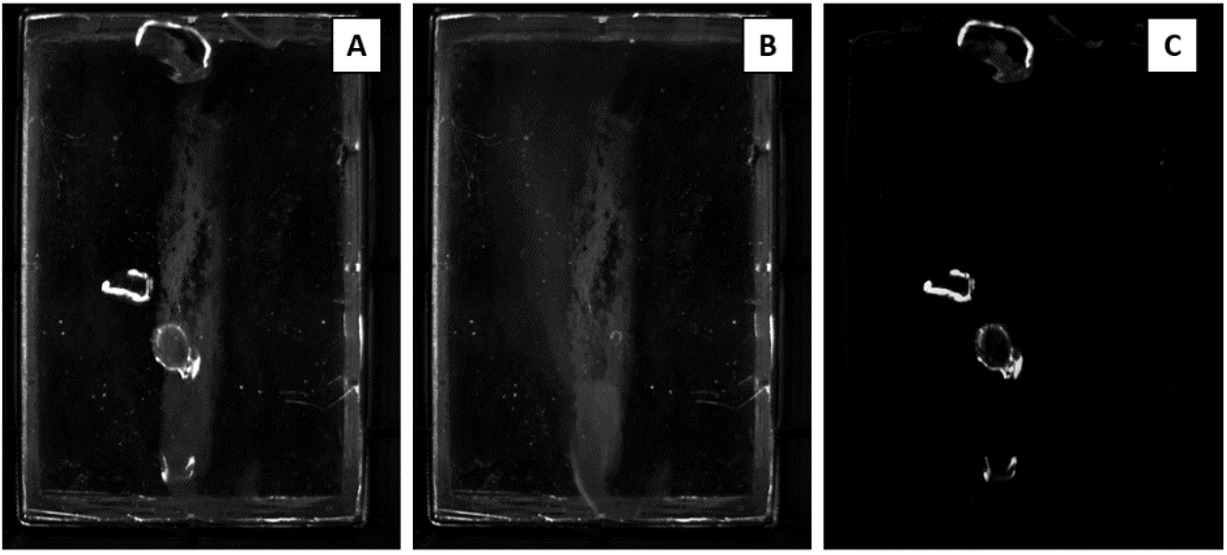}
\caption{Background removal: (A) original image, (B) time-average image, and (C) the result of the mean image subtraction.}
\label{fig:mean_subtraction}
\end{center}
\end{figure}

Before bubble detection, the raw images are pre-processed. The camera introduces a fish-eye lens effect, which is corrected using \textit{OpenCV} following \cite{fisheye-correction} -- the result is shown in Figure \ref{fig:calibration}. Afterward, the images are cropped to the FOV of interest. Next, background is removed from the resulting images by subtracting the mean image over time for an image sequence from all its images (Figure  \ref{fig:mean_subtraction}). This method isolates moving/dynamic image elements from static background, which greatly simplifies further processing. It should be noted that the background is not strictly static. Acid droplets occasionally raise to the surface, so there is a transition between two static background scenes. This can be mitigated by replacing the global mean subtraction with mean background corrections for each static background interval in time.

\clearpage

\begin{figure}[htbp]
\begin{center}
\includegraphics[width=0.85\linewidth]{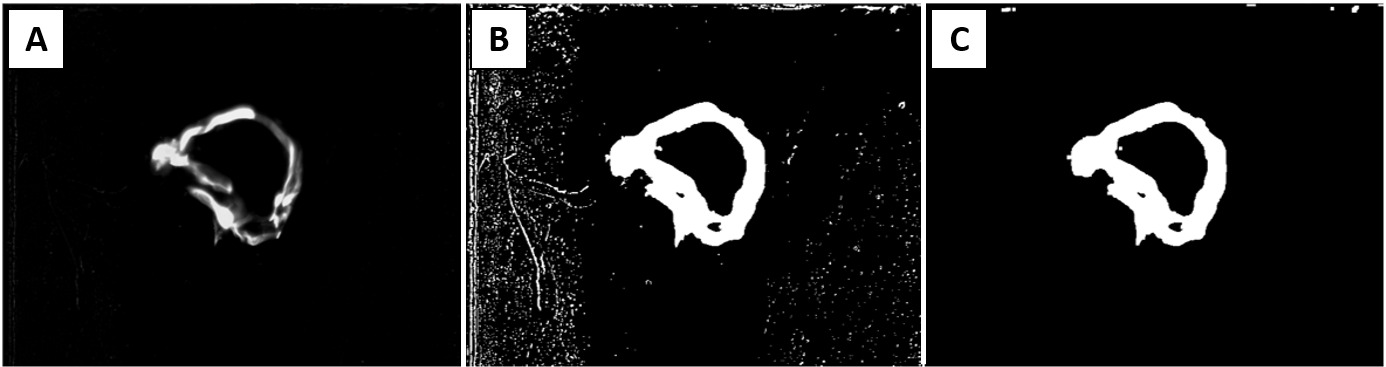}
\caption{Bubble segmentation: (A) original image, (B) manual thresholding, (C) artifact removal using morphological erosion and dilation.}
\label{fig:binarization_artifact_removal}
\end{center}
\end{figure}

The binarization stage helps to outline bubbles more clearly (Figure \ref{fig:binarization_artifact_removal}B). The binarization threshold is manually set to $\sim 4.5 \cdot 10^{-2}$ for all image sequences. It is so, because after mean subtraction wall-bound bubbles are not very bright on the interior, and we should take any advantage to detect their interface. Artifacts are present after binarization, and are removed using morphological erosion \cite{images-mathematical-morphology} and dilation \cite{images-mathematical-morphology} (Figure \ref{fig:binarization_artifact_removal}C). Square kernels are used for both operations. Outlines of bubbles are extracted via a contour recognition algorithm (\textit{OpenCV}) \cite{opencv-contour-recognition}. Contours with areas below a threshold and contours near the FOV edges are removed from analysis.

\clearpage

\subsection{Bubble trajectory recovery}

\subsubsection{Graph-based framework}

\begin{figure}[htbp]
\begin{center}
\includegraphics[width=.95\linewidth]{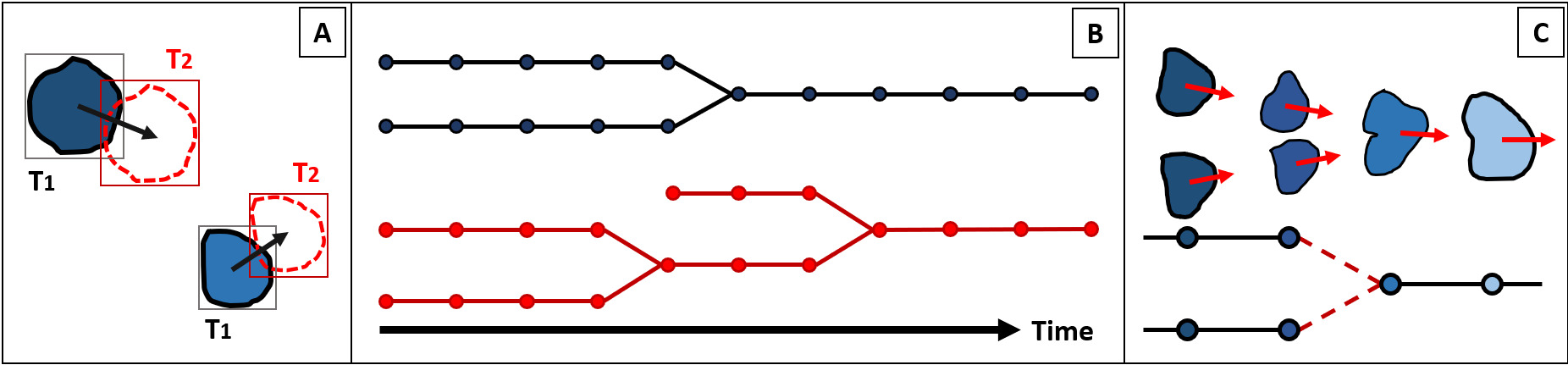}
\caption{(A) Overlaps of bubble contour bounding boxes for two consecutive frames, (B) an example of a graph composed of all two-frame bubble overlap relations, with two disjoint sets of trajectory families, (C) visual and graph representations of a bubble merge event.}
\label{fig:graphs_00a}
\end{center}
\end{figure}

To reconstruct bubble motion, we first analyze bubble positions for all time-adjacent frame pairs. The goal is to determine the relations between bubble contours, which can be established using spatial proximity. High temporal resolution allows only checking for overlap between contours (their bounding boxes) in two sequential frames, as bubbles move only a fraction of their equivalent diameter per frame (Figure \ref{fig:graphs_00a}A). This analysis is performed for all frame pairs, and relations are aggregated into a single \textit{relation graph} as nodes (identified contours) and edges (proximity overlaps).

This framework allows extracting split/merge events naturally, using the notion of nodes' degrees. The bounding box intersection approach overestimates proximity, but it is a safeguard for cases when bubble contours given by segmented reflections in the FOV are fragmented (explained later). By analyzing the relation graph, we can detect disjoint clusters of nodes (Figure \ref{fig:graphs_00a}B). Such clusters represent collections of bubbles, \textit{families}, that only interact with other bubbles within this family. Each family is then analyzed separately with a refined proximity threshold.

\subsubsection{Types of graph structures}

If only translucent type bubbles were present in the data (Figure \ref{fig:exp_illum1}A), the relation graph would contain two types of structures (Figure \ref{fig:graphs_00a}B,C): nodes connected in prolonged \textit{chains}, which represent bubbles that ascend without interacting with other bubbles and without interruptions in trajectories; structures where multiple chains merge or split with/into other chains, which may be attributed to actual bubble split/merge events.

Realistically, however, there are many cases where wall-bound bubbles are not illuminated fully due to chaotic oscillations of their interfaces, so an ideal ring-like reflection is absent. In these cases, binarized images show reflection segments that are visually broken up into multiple fragments, which in the relation graph can be falsely interpreted as split/merge events. In some instances, fragmented contours can be joined using erosion/dilation operations, but only if the fragments are relatively close, which is not always true. In a graph, such false positive events (\textit{optical splits}) are represented by structures where a node chain experiences a brief "explosion" into multiple nodes per time stamp and then "collapses" back into a single chain (Figure \ref{fig:graphs_01}A).

Overall, the resulting graph is a collection of node chains connected by two types of events -- intervals where optical splits/merges of bubbles occur and intervals of bubble proximity within a distance threshold. By examining the structure of the graph, one can differentiate between different types of events and eliminate and/or recover from false positives. This can be done by analyzing connectivity between chain-like graph structures, specifically the number of chains (or \textit{branches}) coming in and going out of an event connecting them. For example, cases with 1 incoming and 1 outgoing (one-to-one) connections are false positives (Figure \ref{fig:graphs_01}A); many-to-one and vice versa connections are merges and splits, respectively  (Figure \ref{fig:graphs_00a}C); many-to-many connections are "mixed" type events, which might indicate complex interaction patterns, or simply a result of multiple bubbles passing close by. It therefore makes sense to classify basic cases and devise a corresponding methodology for graph event recovery/refinement.

\clearpage

\subsubsection{Resolving graph chain connections}

\begin{figure}[bth]
    \begin{center}
        \includegraphics[width=.95\linewidth]{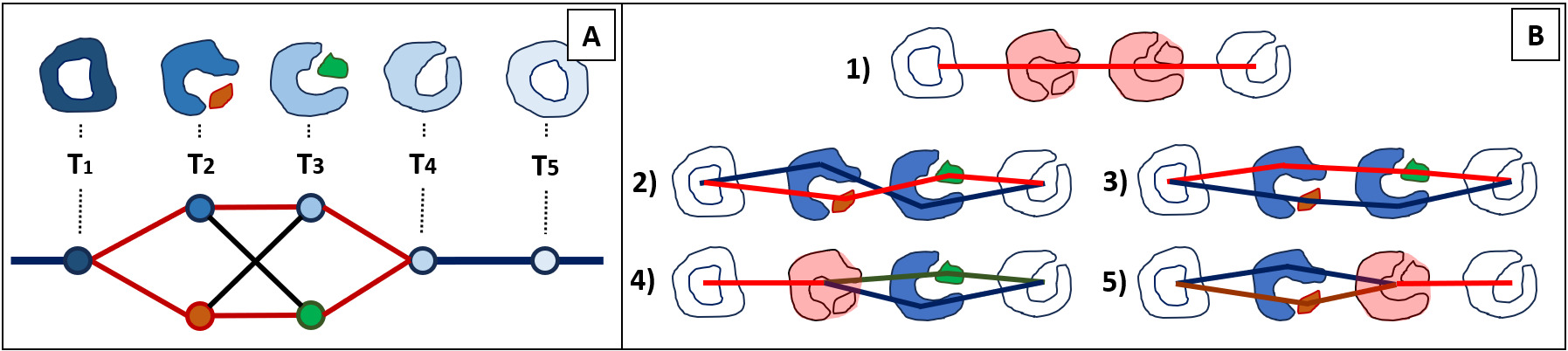}
        \caption{An example of an optical (false positive) bubble split/merge: (A) visual and graph representation of the event, (B) possible sequences of segment connections (or time-evolutions of segments) through the event. Red overlays represent bubbles constructed from all segments (an entire node cluster) at a given time step. Colored lines represent the choices of cluster partitions (evolution paths) selected at each time step. For instance, in (2), the blue line shows a bubble evolving through the event via only the larger blue contours. Timestamps are denoted as $T_k$.}
        \label{fig:graphs_01}
    \end{center}
\end{figure}

\textbf{\textlabel{Case (1).}{case:one}} A false positive event represents a perceived breakup of a single bubble, which is actually an optical artifact. This case is the simplest to detect, as it is represented by a graph event where two chain structures are connected by a one-to-one event. The naive solution is to combine a cluster of nodes at each time step into composite objects, and treat them as missing stages of a longer single-bubble detection event chain. This does not account for cluster elements potentially being visual artifacts. A more rigorous approach is to consider different permutations of elements (i.e., removing/keeping segments in each instance) in a cluster at each time step. This implies that we have to generate different feasible time-evolutions of clusters (Figure \ref{fig:graphs_01}B).

One of the generated evolutions is most physically probable, which can be determined by minimizing a physics-based criterion associated with a given evolution. Here, the criteria are based on conservation of mass and momentum (i.e., as in \cite{mht-x-og}). For the imaged quasi-2D flow, conservation of mass is strongly associated with the conservation of bubble area (convex hull area for clusters), and conservation of momentum is indirectly enforced with a trajectory smoothness criterion. A 2D implementation of momentum conservation is not used for clusters because different permutations of cluster elements can lead to similar centroids, and a similar problem arises for area calculation when contour hulls are used. The criteria minimized for clusters are: the sum of centroid displacements (a natural way to enforce the shortest trajectory paths); the sum of relative bubble area changes between adjacent frames (avoids evolutions with rapid/unphysical bubble area changes); the sum of errors in distances from predicted (explained below) position, and a similar approach is used for expected bubble areas.

Estimated bubble positions and area values for the last criterion are extracted by interpolating prior and posterior history of an event. One single-bubble node chain can be split with fake events multiple times, which in the graph looks like a sequence of chain structures separated by one-to-one events. By combining the resolved states of a larger chain, we can gather longer trajectory history and produce higher-quality predicted missing data.

One concern is whether one can miss a true split-to-merge event inside this event, i.e. if it occurs as a sub-resolution (temporal) event. The current setup has a temporal resolution much higher than expected/observed physical process time scales. If bubbles produced in a split event were stable enough to exist for multiple time steps, they are resolved as separate chain structures in the graph, so the one-to-one event constraint is violated.

\begin{figure}[bth]
    \begin{center}
        \includegraphics[width=0.90\linewidth]{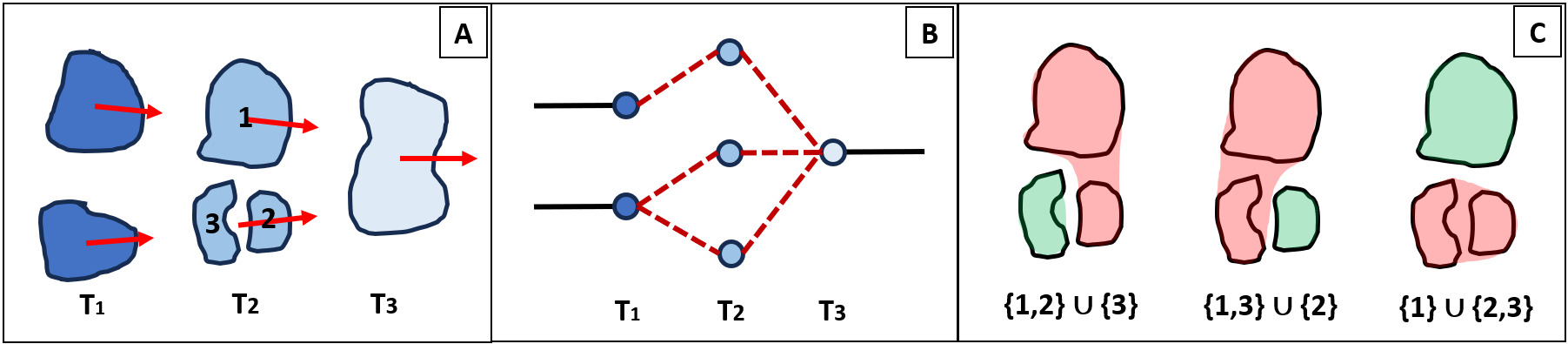}
        \caption{An example of a generalized merge event treatment: (A) visual representation, (B) graph representation, (C) examples of redistribution of cluster elements for branches of a merge event. Timestamps are denoted as $T_k$.}
        \label{fig:graphs_02}
    \end{center}
\end{figure}

\textbf{\textlabel{Case (2a).}{case:two-a}} Events within the proximity threshold are more challenging to define/detect from the graph. High temporal resolution allows us to classify merges/splits as many-to-one/one-to-many events in which multiple branches are involved. The proximity thresholding approach overestimates the distance at which bubbles begin to interact, so split/merge events can be detected prematurely. In practice, transition between graph branches and splits/merges is not clear-cut, an event may take many time steps. If resolving an event requires more than event classification, analysis below the proximity threshold is necessary, and one must refine (i.e., extend) incoming/outgoing bubble paths (branches) as close to the merge/split event as possible. Similarly to \ref{case:one}, one should explore different combinations of cluster elements for each time step. In fact, some of the merging/splitting bubbles may be optical splits (\ref{case:one}), so the method should be generalized to incorporate these cases as well (Figure\ref{fig:graphs_02}A). Unlike \ref{case:one}, where all options are considered simultaneously, here the process should be executed by means of iterative extrapolation (details in \ref{appendix:C}), from the last resolved bubble position in a branch, and its prior (or posterior) history. Each branch ends at a node in a graph where multiple options for further structure selection are present (Figure \ref{fig:graphs_02}B). It is likely that other branches are competing for the same nodes, so the goal is to generate all possible ways to redistribute nodes between competing branches (Figure \ref{fig:graphs_02}C). To redistribute conservatively, it is enough to generate all possible node partitions. The best choice of partitioning, similarly to \ref{case:one}, is determined by minimizing the area and displacement criteria, which in this case are derived from extrapolated branch history.

\begin{figure}[bth]
    \begin{center}
        \includegraphics[width=0.825\linewidth]{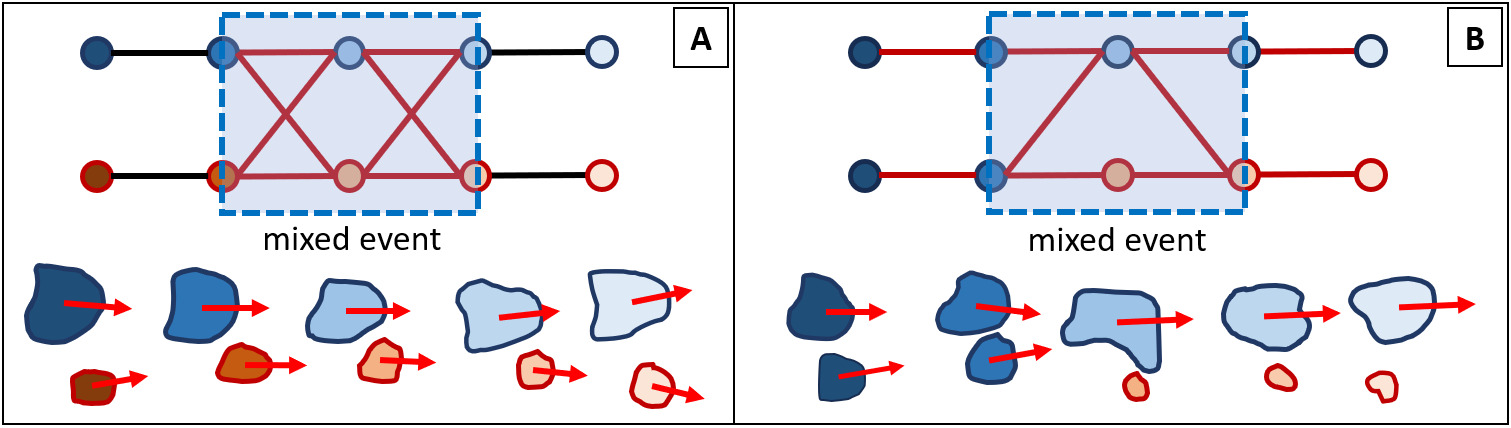}
        \caption{The ambiguity of resolving mixed type events. (A) A case of two bubble passing close by, and (B), the case of a simultaneous sub-resolution (temporal) merge and split.}
        \label{fig:graphs_03}
    \end{center}
\end{figure}

\textbf{\textlabel{Case (2b).}{case:two-b}} Events within the proximity threshold. Mixed merge/split events are the logical extensions of their regular variants as many-to-many events. The nature of these events cannot be deduced easily, since they may be a result of a close passage of two or more bubbles (Figure \ref{fig:graphs_03}A), but a multi-split/merge event is also possible (Figure \ref{fig:graphs_03}B). Experimental observations suggest that close passes-by are more likely, so this case is treated as a regular split/merge using trajectory extrapolation. To resolve possible passes-by, we check whether trajectory extrapolation reaches the outgoing branches.

\begin{figure}[bth]
    \begin{center}
       \includegraphics[width=0.85\linewidth]{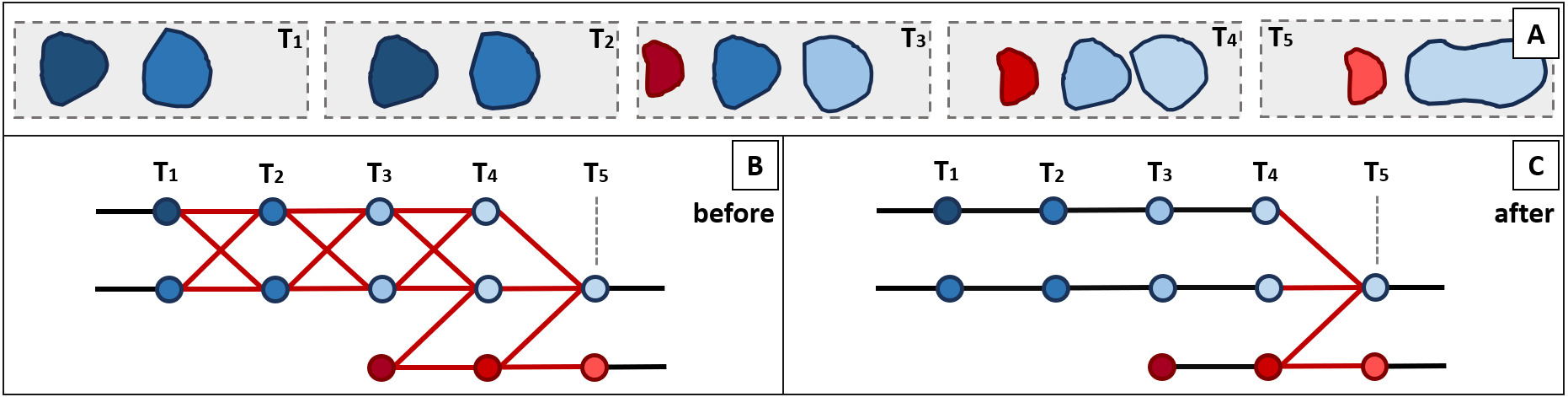}
       \caption{An example of a generalized mixed event treatment. (A) A case of merge of two bubbles with an additional bubble joining the event from out of camera FOV, (B) graph representation of this event, and (C) event graph after one iteration of branch extension.}
       \label{fig:graphs_04}
    \end{center}
\end{figure}

Mixed events may contain within it trajectories of bubbles that were not explicitly resolved. This can happen if the trajectory is too short to be classified as a separate event branch. Often this is the case if the event occurs near the inlet.
%Mixed events may contain implicit (hidden) branches, which are chain structures in the graphs that are too short to be classified as branches. For example, it may be the event that happens near the boundary of an image, and incoming bubbles did not accumulate sufficiently long history. 
An example can be seen on Figure \ref{fig:graphs_04}A, which shows the interaction of three bubbles, where only two of them register as explicit incoming branches (Figure \ref{fig:graphs_04}B), and only two outgoing branches present. This configuration of "2 in"-"2 out" event hides the presence of internal merge and disguises the event as a regular bubble pass-by. After using the branch extension algorithm, we successfully resolve the time steps up to a merge node, but the hidden branch stays connected to this event. However, it is now resolved as an additional incoming branch. To properly resolve this and similar cases, event processing must be performed multiple times, achieving refined results with each iteration.

\clearpage

\section{Results}

Using the developed optical imaging setup, a comprehensive dataset was acquired, spanning a range of flow rates for different MF configurations (Table \ref{tab:experiment-summary}). The utilized MF systems are the same as in \cite{birjukovsArgonBubbleFlow2020, birjukovsPhaseBoundaryDynamics2020, birjukovs2021resolving}, and produce a rather homogeneous MF within the metal vessel. We \textit{assume} that bubble recognition in our optical experiments is more error-prone/difficult than for x-ray/neutron radiography. To mitigate this and to increase statistical significance of observed physics, we perform multiple imaging series per MF/flow rate instance. The scope of this paper is the newly developed optical imaging methodology and the associated data analysis, plus their capability demonstration. A full physical analysis of the entire accumulated image dataset is therefore reserved for a follow-up paper. Here, we present some of the quali-/quantitative bubble flow characteristics that can be derived from the acquired images. As an example, we take one series for every flow rate for the following MF configurations: no MF (\textit{Field OFF}), $B=125~mT$ horizontal field system (HFS), $B=125~mT$ vertical field system (VFS), and $B=200~mT$ HFS. Note that every image sequence has $15\text{K}$ images, which at $350$ FPS is $\approx 42.85~s$ of flow time.

\begin{table}[h!]
\centering
\resizebox{0.775\textwidth}{!}{
 \begin{tabular}{|c c c c|} 
 \hline
 MF system & MF induction $B$, $mT$ & Imaging series & Gas flow rate, $sccm$ 
\\
 \hline
 \hline
 Field OFF & 0 & 7 & 100, 150, 200, 250, 300, 350, 400 \\ 
 HFS & 125 & 5 & 100, 150, 200, 250, 300, 350, 400  \\
 HFS & 200 & 5 & 100, 150, 200, 250, 300, 350, 400  \\
 HFS & 265 & 5 & 100, 150, 200, 250, 300, 350, 400  \\
 VFS & 125 & 5 & 100, 150, 200, 250, 300, 350, 400  \\
 \hline
 \end{tabular}
 }
 \caption{A summary of the performed optical imaging experiments. Here, V-/HFS stand for vertical and horizontal MF systems, respectively.}
 \label{tab:experiment-summary}
\end{table}

\begin{figure}[H]
    \begin{center}
        \includegraphics[width=0.875\linewidth]{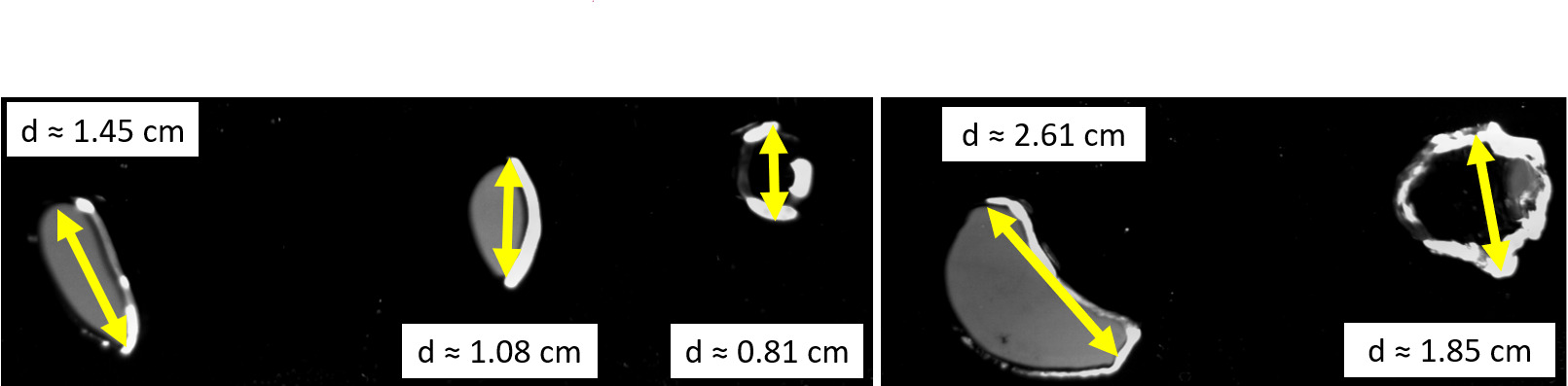}
        \caption{A characteristic example of how bubbles present in captured images, with highlighted bubble sizes.}
        \label{fig:bubble-size-1}
    \end{center}
\end{figure}

\begin{figure}[H]
\begin{center}
\includegraphics[width=0.47\linewidth]{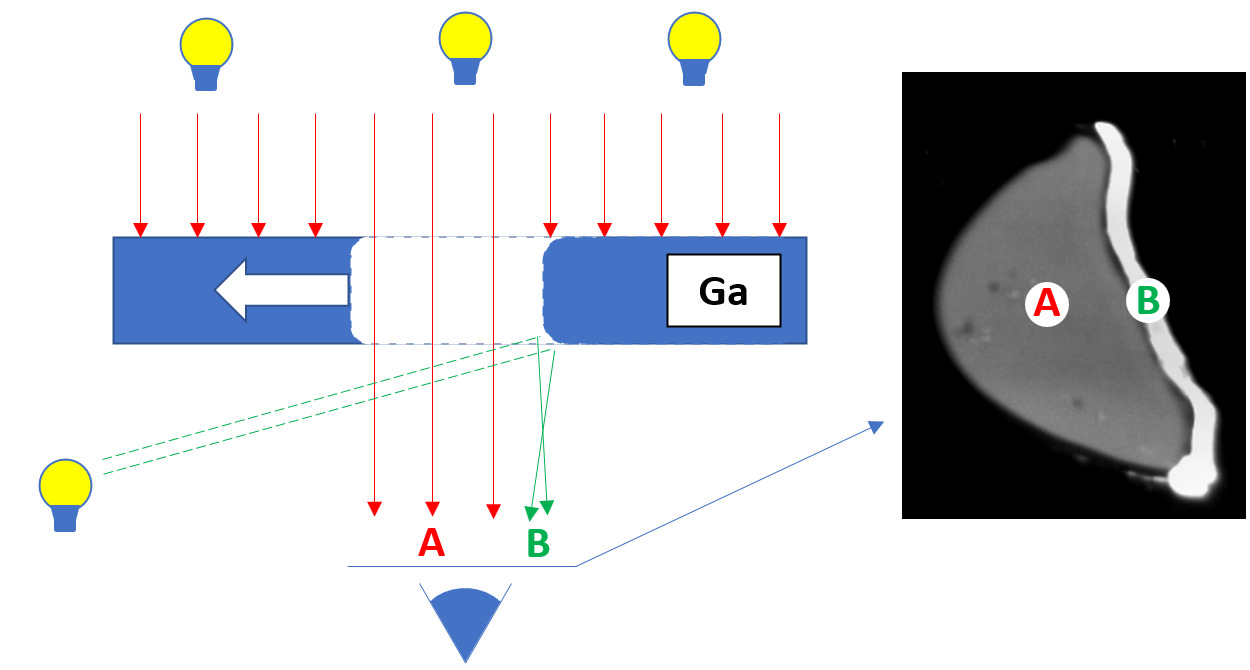}
\caption{A schematic illustration of how a bubble meniscus reflection forms in images due to the side illumination.}
\label{fig:side-illumination-meniscus}
\end{center}
\end{figure}

\begin{figure}[htbp]
    \begin{center}
        \includegraphics[width=0.815\linewidth]{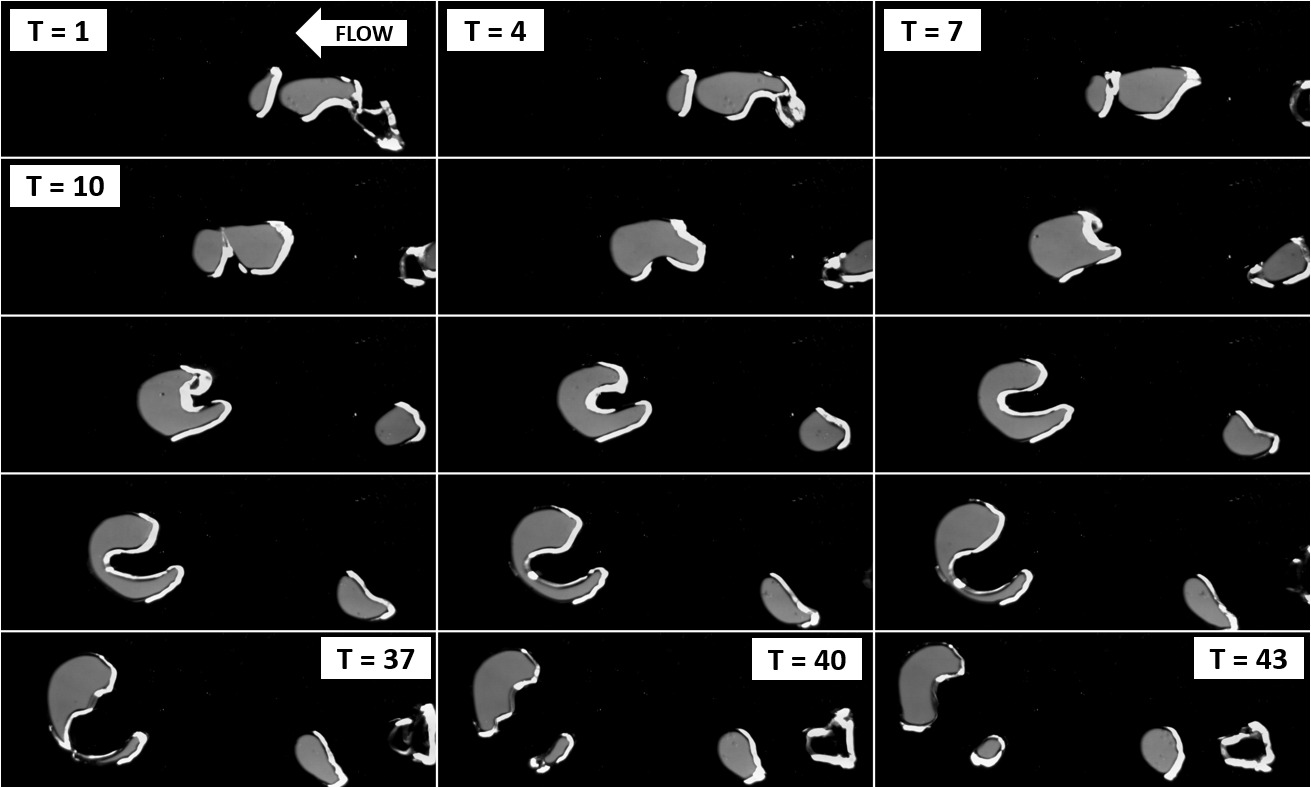}
        \caption{A bubble merge event (T = 1-7) followed by a split event (T = 37-40), where T is the relative frame number.}
        \label{fig:bubble-dynamics-example}
    \end{center}
\end{figure}

\begin{figure}[htbp]
    \begin{center}
        \includegraphics[width=0.90\linewidth]{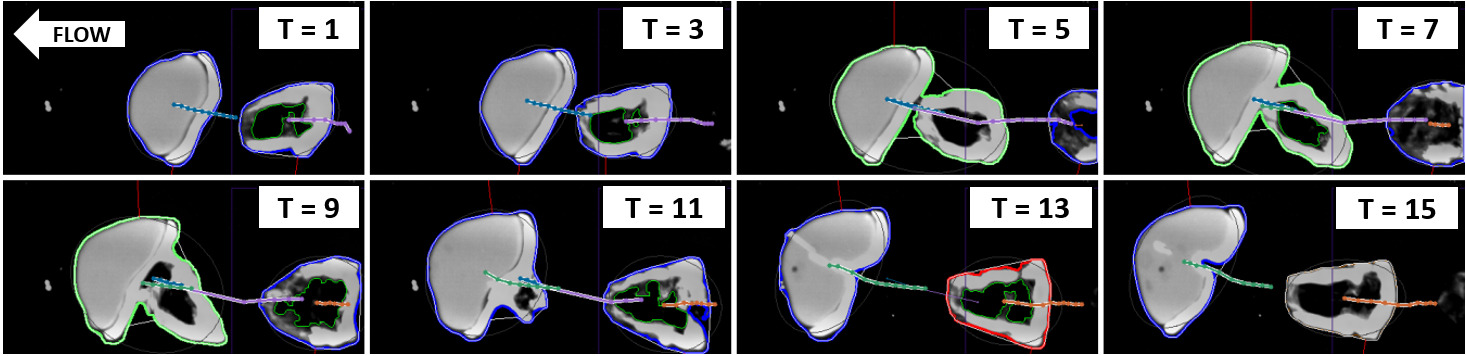}
        \caption{A merging event involving a translucent and a wall-bound bubble, with highlighted bubble contours and trajectories. T is the relative frame number.}
        \label{fig:bubble-merge-example}
    \end{center}
\end{figure}

Figures \ref{fig:bubble-size-1} and \ref{fig:bubble-dynamics-example}-\ref{fig:bubble-merge-example} showcase the typically observed bubbles (translucent and highlighted via the side illumination) and their dynamics. Note the brighter spots/lines at the bottom of the translucent (gray interior) bubbles. These are the reflections formed due to the side illumination over the metal meniscus, forming in the Hele-Shaw cell at the trailing bubble edge, as illustrated in Figure \ref{fig:side-illumination-meniscus}. This reveals bubble shape/area that, unlike in X-ray/neutron imaging, are otherwise lost. However, the top edge, essentially an inverted meniscus, is rarely illuminated (an example of partial illumination shown in Figure \ref{fig:bubble-size-1}), which must be considered when analyzing data. However, one can likely assume that geometric confinement in the metal cell and the surface tension for GaInSn and argon are sufficient to produce a symmetric high-curvature meniscus (inverted at the top).

\clearpage

\begin{figure}[htbp]
\begin{center}
\includegraphics[width=0.6725\linewidth]{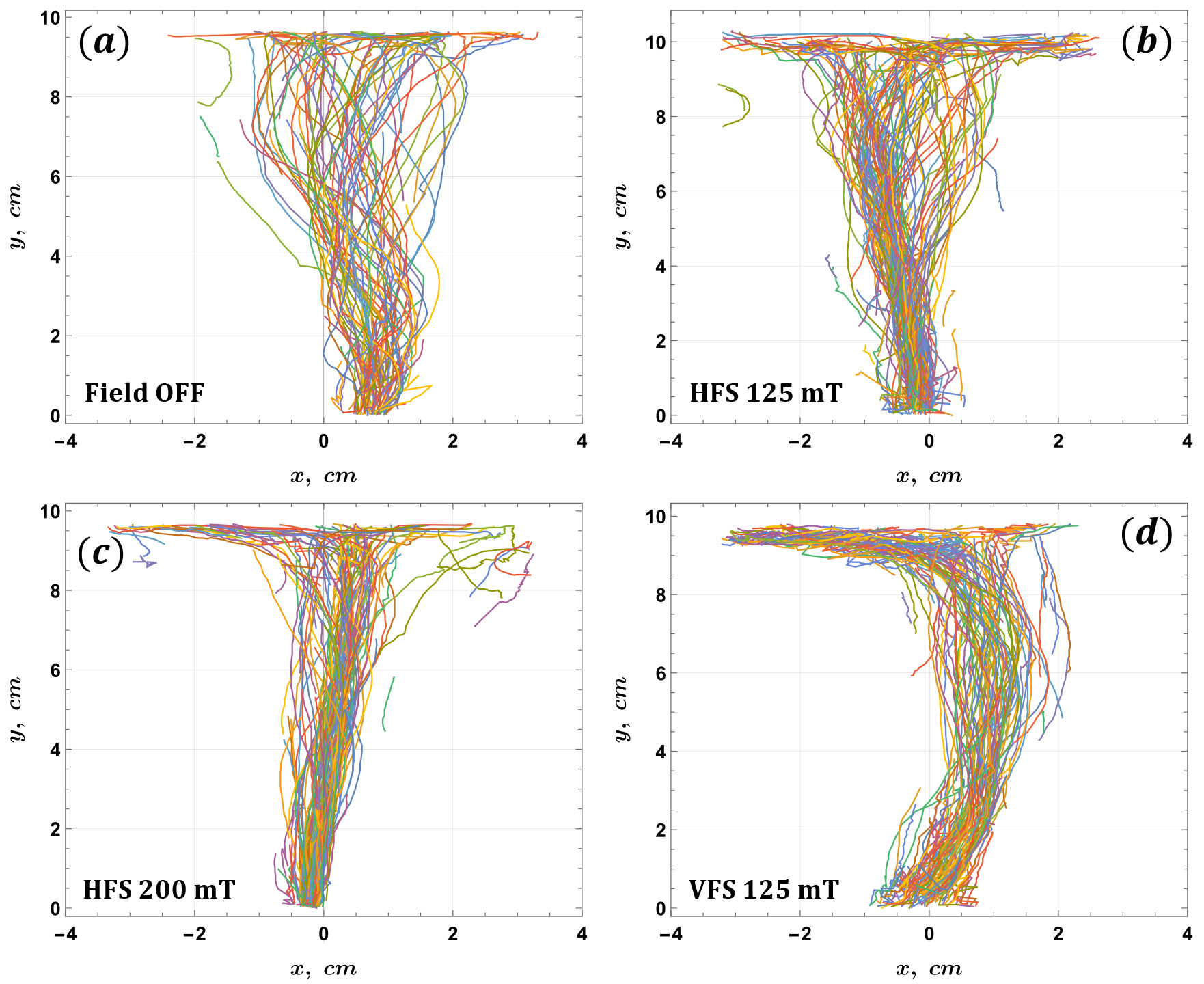}
\caption{An example of bubble trajectories ($>10$ nodes long, color coded by their IDs) observed for four different MF configurations at $350~sccm$. Here $x$ and $y$ are coordinates along the metal cell width and height, respectively.}
\label{fig:example-trajectories}
\end{center}
\end{figure}

Importantly, as seen in Figure \ref{fig:bubble-merge-example}, the developed image processing code can track both translucent and side-illuminated bubbles, as well as resolve their interactions, such as splits and merges. With trajectories reconstructed with the methods explained above, one can not only visualize them (Figure \ref{fig:example-trajectories}), but also perform a more rigorous analysis.

\begin{figure}[H]
\begin{center}
\includegraphics[width=0.76\linewidth]{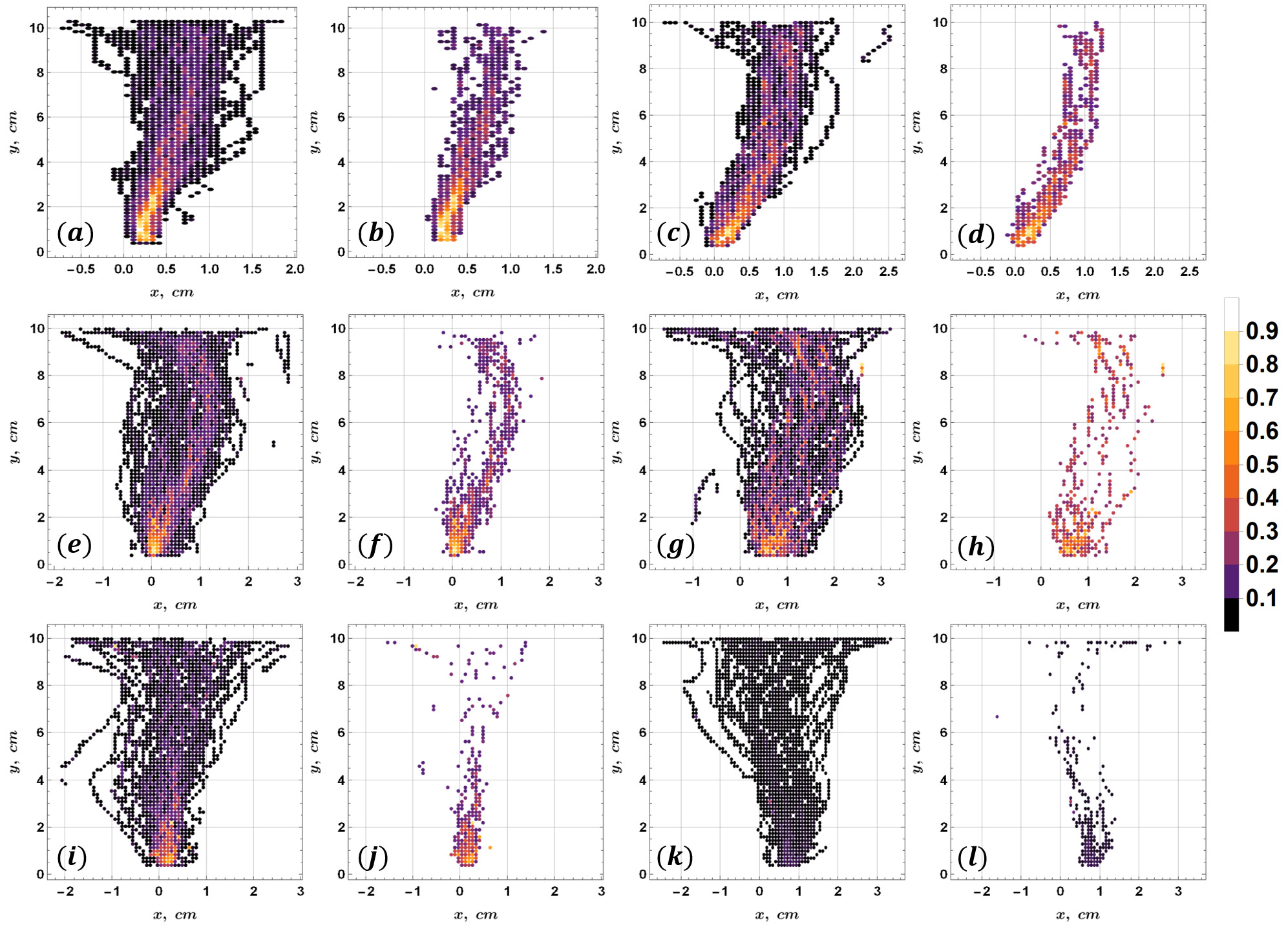}
\caption{Normalized bubble detection density histograms for image sequences without applied MF. The flow rates are: (a-b) $100~sccm$, (c-d) $150~sccm$, (e-f) $200~sccm$, (g-h) $250~sccm$, (i-j) $300~sccm$, (k-l) $350~sccm$. Bin size is $(dx,dy) = (0.75,1.5)~ mm$, and they are color-coded by detection counts. Two images are shown per flow rate: one with all bins, another with bins that have 6+ bubble detection instances. Each histogram is normalized separately. Please note that the plot scales are not identical.}
\label{fig:density-bins-field-off}
\end{center}
\end{figure}

\begin{figure}[H]
\begin{center}
\includegraphics[width=0.76\linewidth]{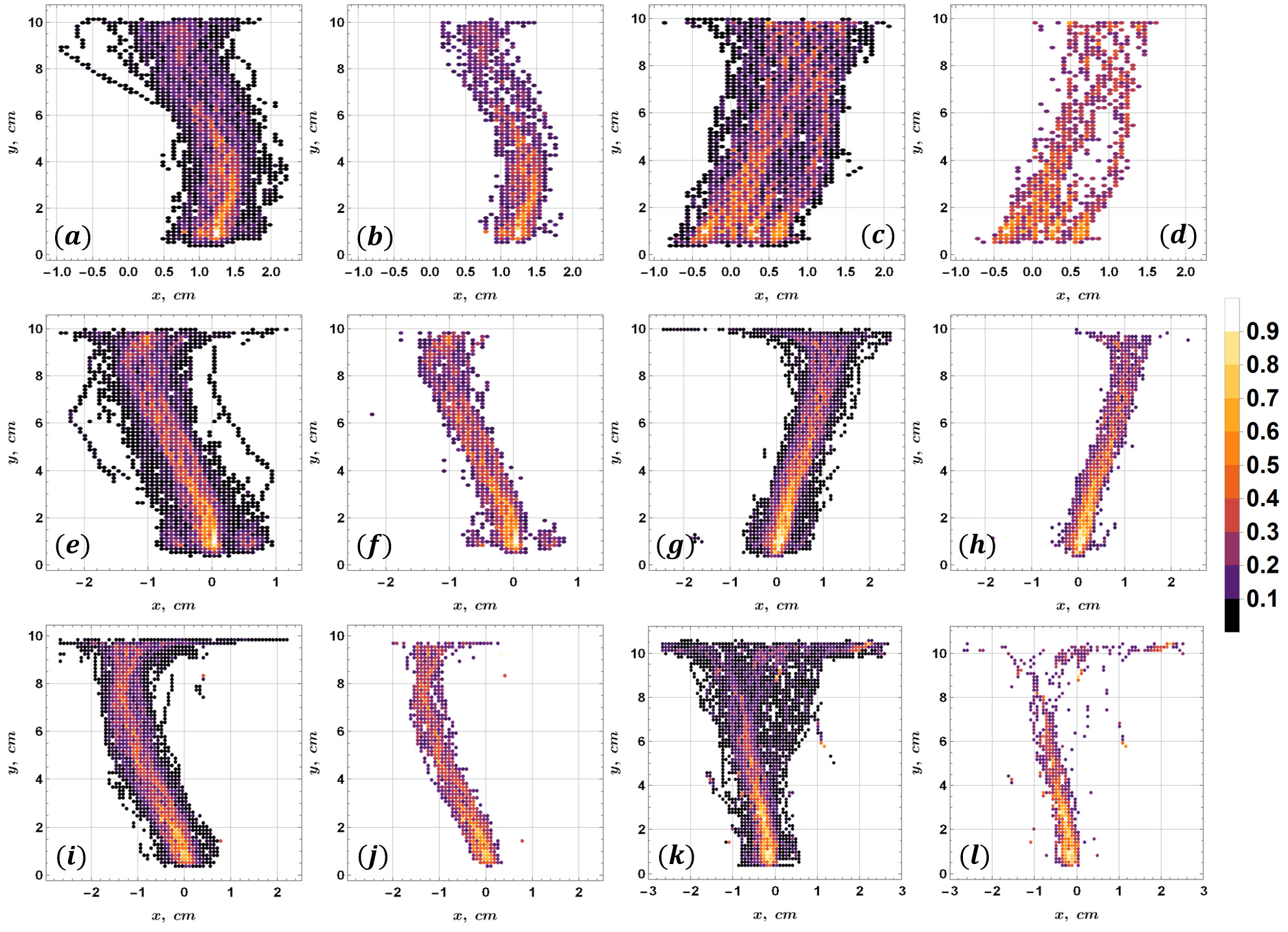}
\caption{Normalized bubble detection density histograms for image sequences with HFS $125~mT$. The flow rates are: (a-b) $100~sccm$, (c-d) $150~sccm$, (e-f) $200~sccm$, (g-h) $250~sccm$, (i-j) $300~sccm$, (k-l) $350~sccm$.}
\label{fig:density-bins-HFS-125mT}
\end{center}
\end{figure}

\begin{figure}[H]
\begin{center}
\includegraphics[width=0.76\linewidth]{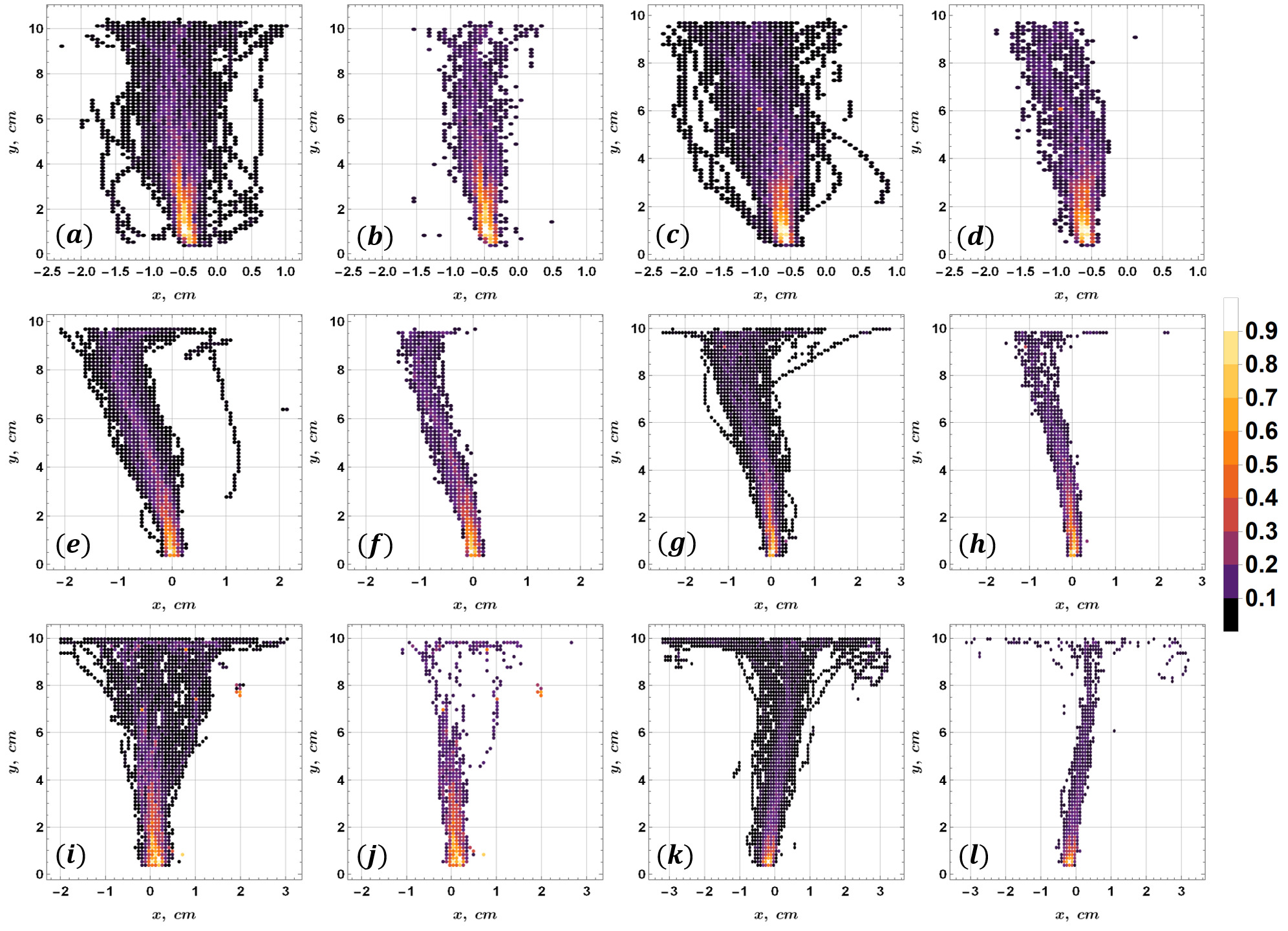}
\caption{Normalized bubble detection density histograms for image sequences with HFS $200~mT$. The flow rates are: (a-b) $100~sccm$, (c-d) $150~sccm$, (e-f) $200~sccm$, (g-h) $250~sccm$, (i-j) $300~sccm$, (k-l) $350~sccm$.}
\label{fig:density-bins-HFS-200mT}
\end{center}
\end{figure}

\begin{figure}[H]
\begin{center}
\includegraphics[width=0.76\linewidth]{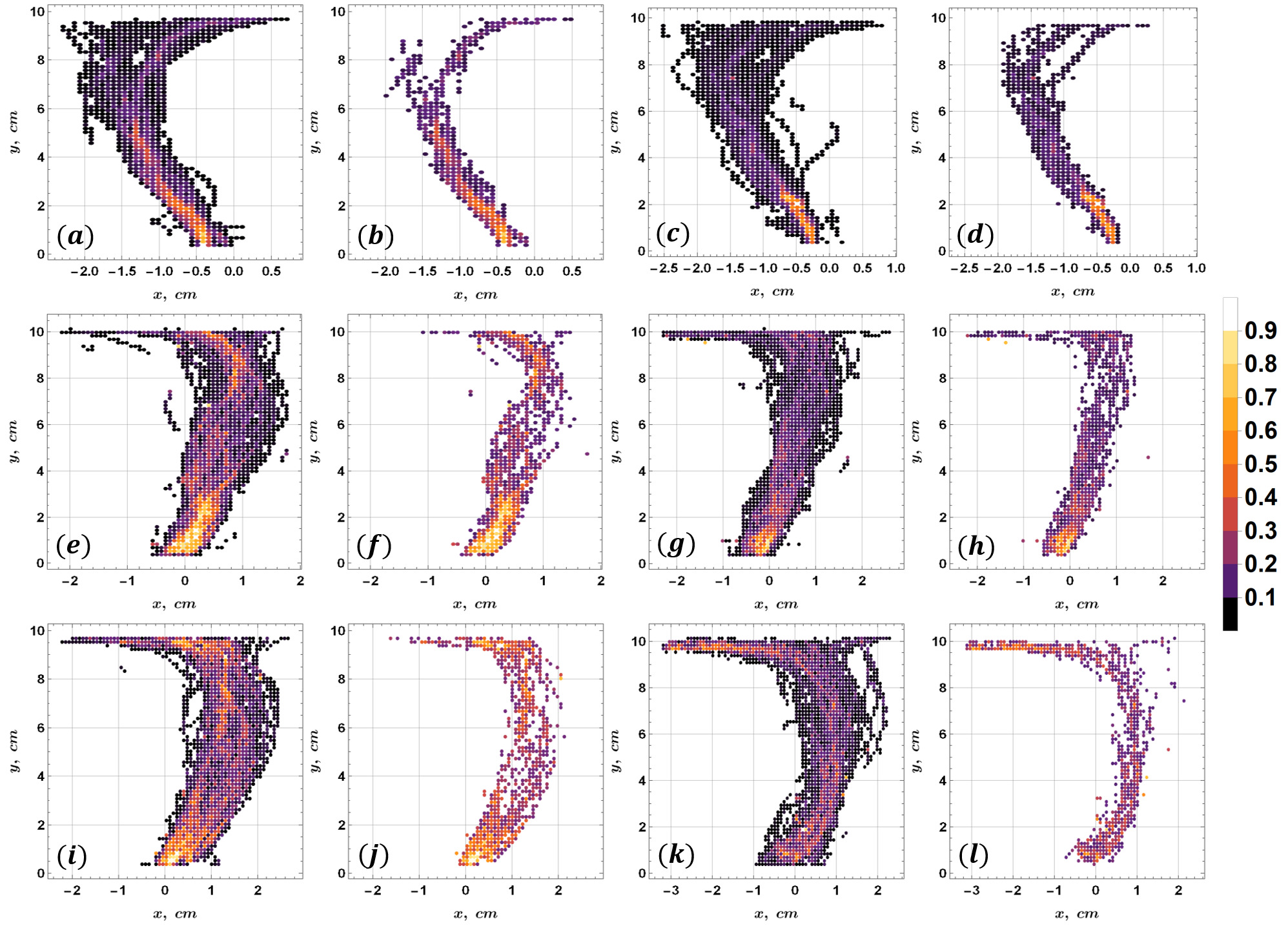}
\caption{Normalized bubble detection density histograms for image sequences with VFS $125~mT$. The flow rates are: (a-b) $100~sccm$, (c-d) $150~sccm$, (e-f) $200~sccm$, (g-h) $250~sccm$, (i-j) $300~sccm$, (k-l) $350~sccm$.}
\label{fig:density-bins-VFS-125mT}
\end{center}
\end{figure}

For instance, one way to describe the effects of flow rate and MF variations is to take all trajectories observed for a given flow rate and MF, and construct a number density histogram. Doing this for all the image sequences considered for this paper yields Figures \ref{fig:density-bins-field-off}-\ref{fig:density-bins-VFS-125mT}. Several observations can be made here. In Figure \ref{fig:density-bins-field-off}, which corresponds to the case with no applied MF, one can see that, as the flow rate is increased, the initially clear and coherent bubble chain shape, especially visible in (b,d,f), starts to disappear as the flow becomes more disordered. The disorder onset can be attributed to wake flow vortex detachment \cite{hele-shaw-bubble-trajectory-instability, prl-path-instability,shape-and-wake-simulations,clift-bubbles}, bubble hydrodynamic interactions without direct contact \cite{hele-shaw-bubble-swarm-1, hele-shaw-bubble-swarm-2}, as well as the observed flow and trajectory perturbations caused by bubble merges and splits, which become significantly more frequent at $200~sccm$ and above. It is clearly visible in (g), where, instead of a clear "core" of the chain, one has tendril-like structure. This culminates with (k,l), where no obvious structure is observable at all.

Conversely, when MF is applied, one can see that with $125~mT$ HFS (Figure \ref{fig:density-bins-HFS-125mT}) mean bubble chain shapes are clear throughout the considered flow rate range, with, perhaps, the exception of $350~sscm$ (l), where bubble motion near the metal free surface (at the $\sim 9$-$10~cm$ mark) is less coherent. Note that at higher flow rates there is a tendency for the number density bins to form a rather clear band at the $10~cm$ mark (g,i,k). From experimental observations, this could be due to bubble splitting that occurs more frequently closer to the top of the vessel, where bubbles often travel in opposite horizontal directions. This observation could also result from bubbles that have reached the top not always leaving the metal volume or bursting, but rather delaying at the free surface for a brief time interval. This could prevent trailing bubbles from freely ascending, forcing horizontal deflection. In addition, vortices usually form about the bubble chain, near the upper corners of the metal vessel \cite{hele-shaw-bubble-vortex-interaction, hzdr-bubbles-mf} -- these too could promote horizontal bubble motion of near the free surface, especially they are less disrupted by smaller bubble wake vortices, the latter being damped by MF. Finally, these patterns could also be partially explained by particularly large bubbles partially leaving the FOV at the very top of the cell, and appearing as splitting into horizontally moving components.

The case of $200~mT$ HFS, seen in Figure \ref{fig:density-bins-HFS-200mT}, is somewhat similar to $125~mT$ HFS in that, as one would expect, motion coherency is preserved, to a degree, with increasing flow rate. Here the bubble chain path is very clear, even for $350~sccm$, as seen in (l). It is also worth noting that in (k), and faintly in (l), one can see some bin clusters in the upper-right corner of the plot. These are likely due to some trajectories forming loops, i.e. bubbles being directed back into the metal volume and away from the free surface. This is most likely the effect of vortices that form about the bubble chain \cite{hele-shaw-bubble-vortex-interaction, hzdr-bubbles-mf, hzdr-ibm-bubbles-thesis, hzdr-bubble-plumes-rect-columns, x-ray-bubble-chain-simulate}. It was also observed for both HFS cases that, once the flow rate is set and bubble chain flow is initialized, the bubble chain is often "locked" into a certain shape until perturbed strongly enough due to free surface oscillations or bubble split/merge events. This is because bubbles tend to ascend in a chain by following the paths of the leading bubbles, where the pressure is significantly lower due to the wake flow \cite{hele-shaw-bubbles-vof, hele-shaw-high-re-bubbles, hele-shaw-bubble-vortex-interaction, hzdr-bubbles-mf, hzdr-ibm-bubbles-thesis}.

Interesting trends can be seen in Figure \ref{fig:density-bins-VFS-125mT}, which is the case with $125~mT$ VFS. It could be argued, comparing with Figure \ref{fig:density-bins-HFS-125mT}, that vertical MF of the same strength preserves the coherence of bubble motion in a chain slightly better than horizontal MF. A clear feature that persists for all flow rates is that the bubble chain has a very clear curved shape, distinct from what is observed in Figures \ref{fig:density-bins-field-off}-\ref{fig:density-bins-HFS-200mT} with HFS or without MF. Here, again, chain shape locking is observed consistently, but looks more pronounced. It is likely that a large vortex forms at the chain and further enforces its shape. In addition, the tendency for bubble detection to occur along the near-surface zone is much clearer than with HFS, and is obvious at flow rates above $150~sccm$.

While the aim is to eventually use the density histograms for bubble detections to later describe the underlying physics quantitatively, in their current form they do little to help objectively quantify how MF stabilizes bubble flow and how increasing flow rate disrupts it. However, one can take the sets of all trajectories for each flow rate and MF configuration, and compute the envelopes of all bubble centroids. One can then compute how the width of these envelopes varies over the elevation in the metal cell, which should be a good metric for how MF affects bubble chains.

\begin{figure}[htbp]
\begin{center}
\includegraphics[width=0.7\linewidth]{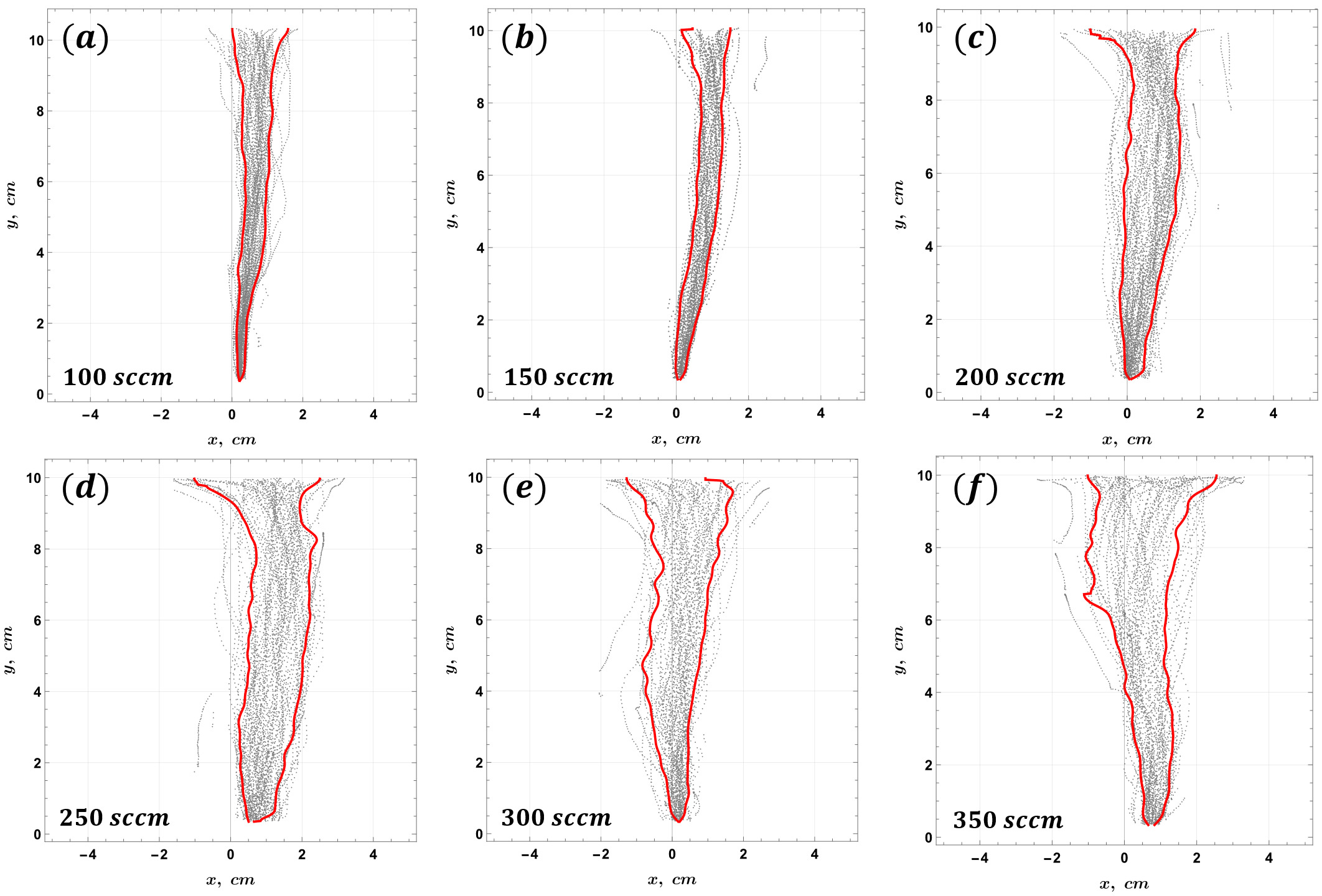}
\caption{Bubble trajectory envelopes for the case without applied MF, computed as QSEs following \cite{antonov-qse, birjukovs2021resolving}. Here, sub-figures (a-f) show envelopes (red contours) over the set of bubble centroids (gray dots) for flow rates $150~sccm$ through $350~sccm$ (Table \ref{tab:experiment-summary}). $x$ and $y$ are coordinates along the metal cell width and height, respectively.}
\label{fig:trajectory-envelope-examples}
\end{center}
\end{figure}

To derive the bubble trajectory/chain envelopes, one can compute quantile spline envelopes (QSE) \cite{antonov-qse} for the set of all bubble centroids. This is done using in \textit{Wolfram Mathematica} using the code (package) available on \textit{GitHub}: \href{https://github.com/antononcube/MathematicaForPrediction/blob/master/QuantileRegression.m}{Anton Antonov
(\textit{antononcube}): MathematicaForPrediction/QuantileRegression.m}. The parameters for QSE generation and regularization are as in \cite{birjukovs2021resolving}. An example of QSEs computed for sets of bubble trajectories is shown in Figure \ref{fig:trajectory-envelope-examples}.

\clearpage

\clearpage

\clearpage

\begin{figure}[htbp]
\begin{center}
\includegraphics[width=0.75\linewidth]{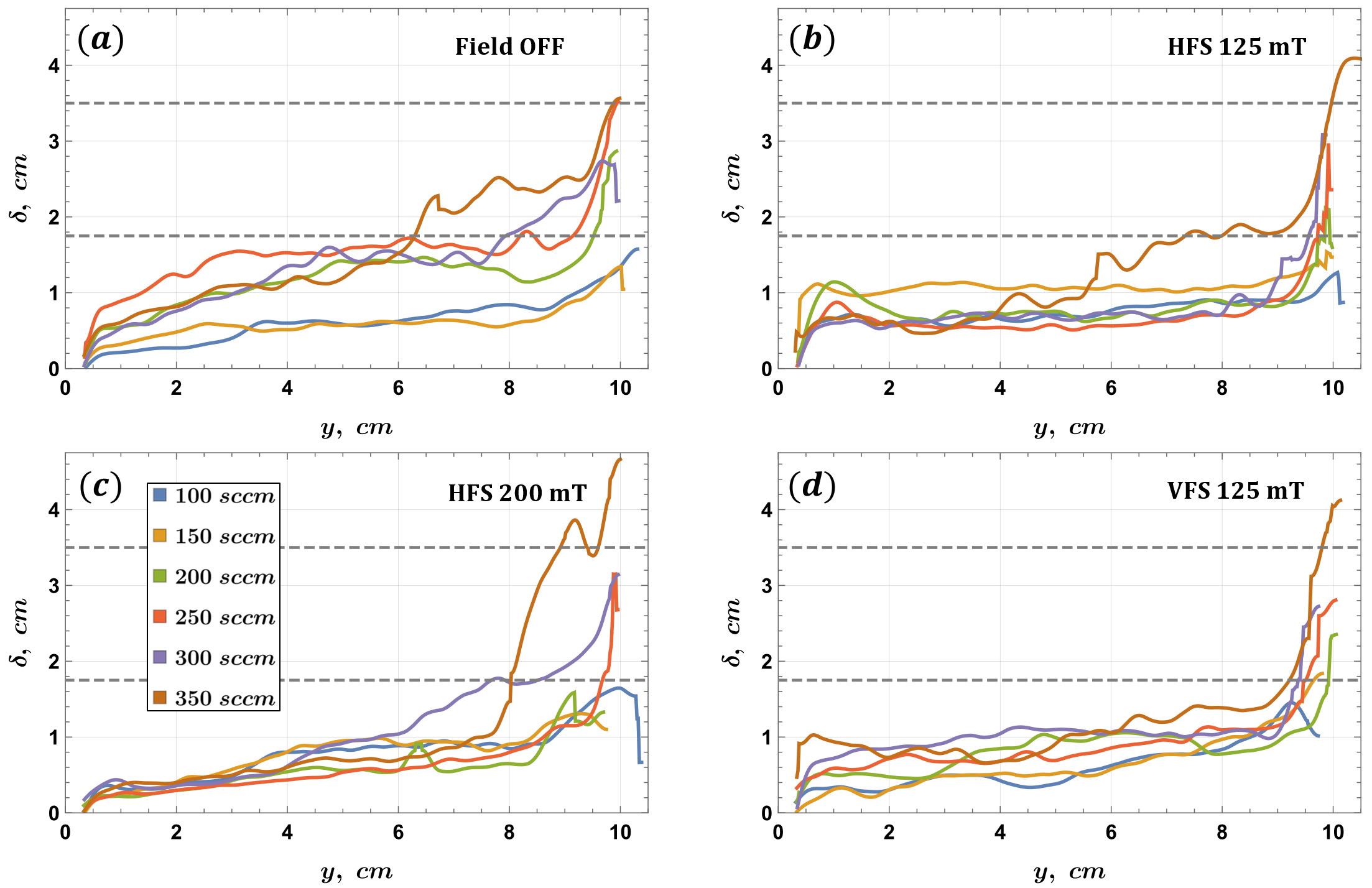}
\caption{Trajectory envelope widths $\delta$ over elevation $y$ from metal cell bottom for flow rates shown in the legend in (c). The four cases are: (a) no applied MF, (b) $125~mT$ HFS, (c) $200~mT$ HFS, (d) $125~mT$ VFS. All plots have identical plot ranges and scales, and the dashed gray lines are for visual reference.}
\label{fig:envelope-width-all-field-configs}
\end{center}
\end{figure}

\begin{figure}[htbp]
\begin{center}
\includegraphics[width=0.75\linewidth]{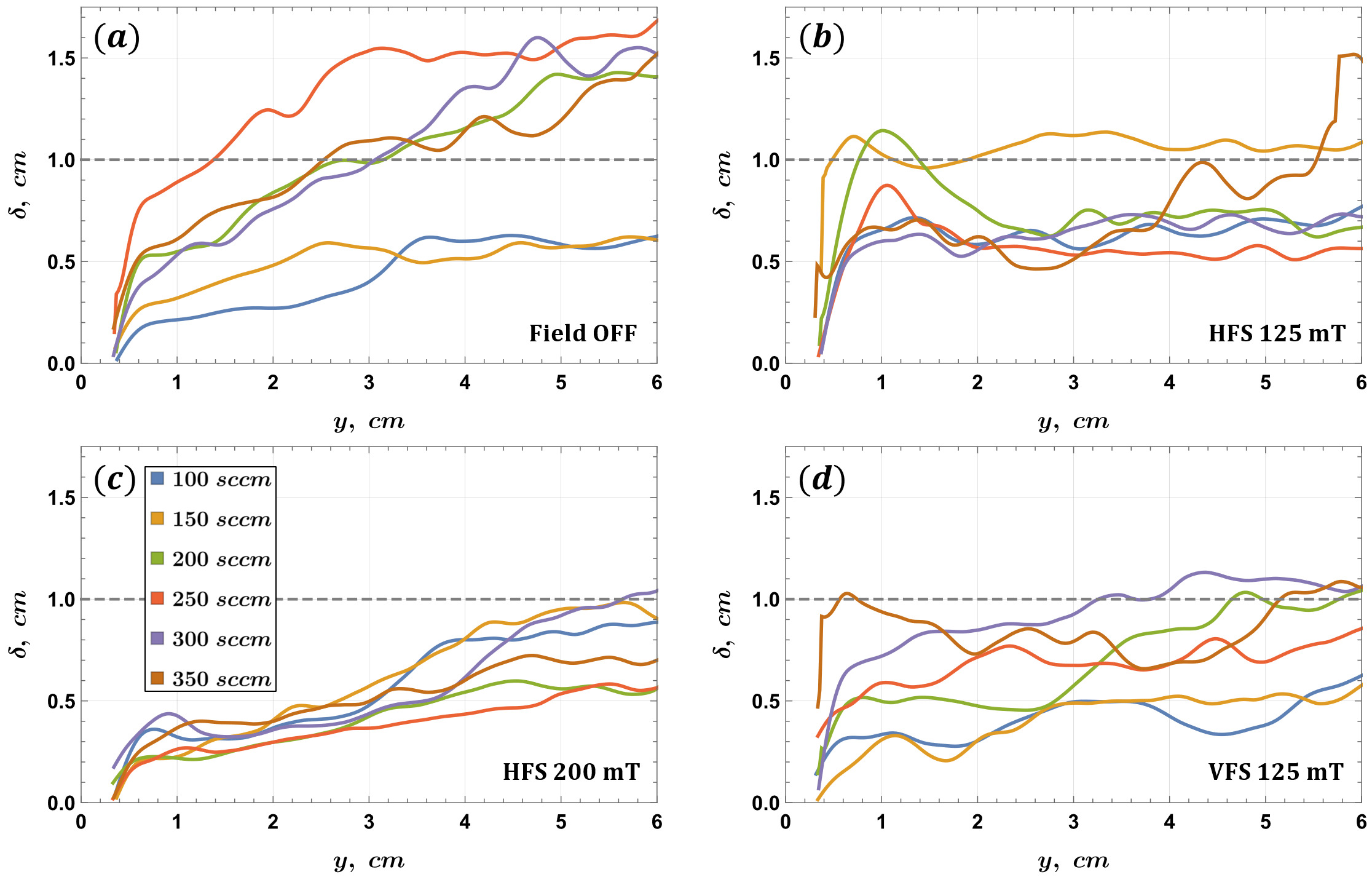}
\caption{Trajectory envelope widths $\delta$ over elevation $y$ from metal cell bottom: (a) no applied MF, (b) $125~mT$ HFS, (c) $200~mT$ HFS, (d) $125~mT$ VFS.}
\label{fig:envelope-width-all-field-configs-zoom}
\end{center}
\end{figure}

Once envelopes and their widths $\delta$ over elevation are computed for all cases, one can get an overview of the effects of varying MF and flow rate, as seen in Figures \ref{fig:envelope-width-all-field-configs}-\ref{fig:envelope-width-all-field-configs-zoom}. The $\delta$ plots clearly show how magnetic field stabilizes bubble chains. For instance, observe the differences between Figures \ref{fig:envelope-width-all-field-configs}a and b -- across the entire flow rate range, the envelopes are significantly thinner for $125~mT$ HFS versus no MF, except near the free surface. Interestingly, while $\delta$ is even smaller for $200~mT$ HFS in the $1$-$6~cm$ range (better seen in Figure \ref{fig:envelope-width-all-field-configs-zoom}), bubble trajectories seem to destabilize outside this range for $300$ and $350~sccm$. However, since we are currently analyzing only 1 image sequence per flow rate, one should keep in mind that this could be not the case in general.

Figure \ref{fig:envelope-width-all-field-configs}d suggests that $125~mT$ VFS is generally slightly better at stabilizing bubble trajectories than HFS of the same strength, and seems comparable to or better than $200~mT$ in the $8$-$10~cm$ range. Looking more closely at the lower half of the vessel height in Figure \ref{fig:envelope-width-all-field-configs-zoom}, one can see that, while for $125~mT$ HFS trajectories are more stable after the $1$-$1.5~cm$ mark, there is a considerable local maximum in $\delta$ within the first $1.5~cm$. For $200~mT$ HFS, this maximum is shifted towards $y=0~cm$ and strongly diminished. Remarkably, this does not manifest at all for $125~mT$ VFS, except at $350~sccm$. VFS also results in more smoothly increasing $\delta$ than with HFS of the same strength.

\begin{figure}[htbp]
\begin{center}
\includegraphics[width=0.6\linewidth]{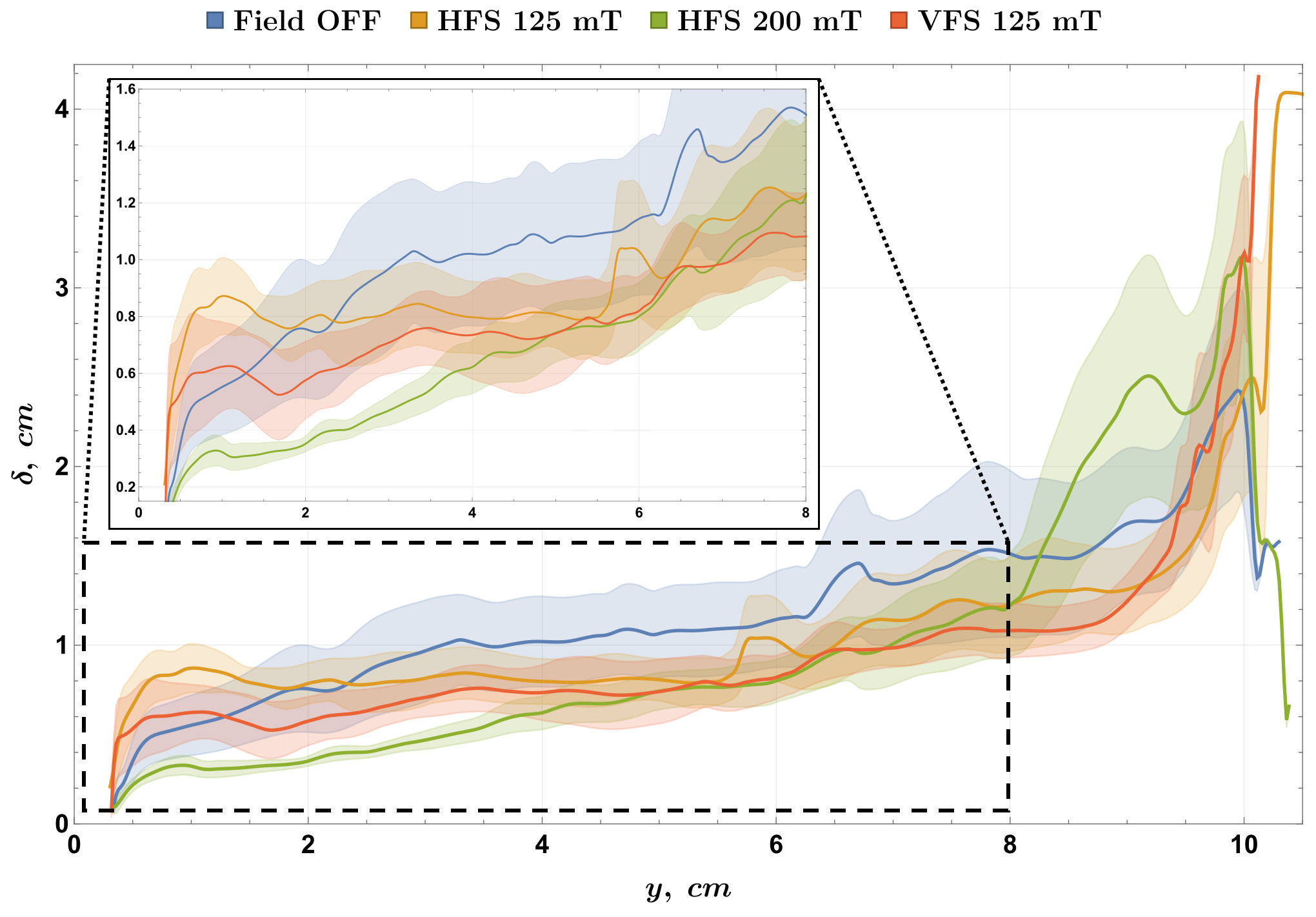}
\caption{Range bands and their middle lines for trajectory envelope width $\delta$ over elevation $y$ for all MF configurations. Note that the bands are shrunk by a factor of $0.25$ for visual clarity.}
\label{fig:envelope-width-bands-all-field-configs}
\end{center}
\end{figure}

For a more general (but "coarser") assessment, one can plot the $\delta$ range bands across all flow rates over elevation, and their middle lines, which is shown in Figure \ref{fig:envelope-width-bands-all-field-configs}. In this representation, it would seem that VFS is indeed more effective at bubble chain/trajectory stabilization than HFS of the same strength, while $200~HFS$ outperforms both in the $0$-$8~cm$ range. As seen in the inset, $125~mT$ HFS/VFS introduce instability in the $0$-$1.5~cm$ range.

\clearpage

\begin{figure}[H]
\begin{center}
\includegraphics[width=0.735\linewidth]{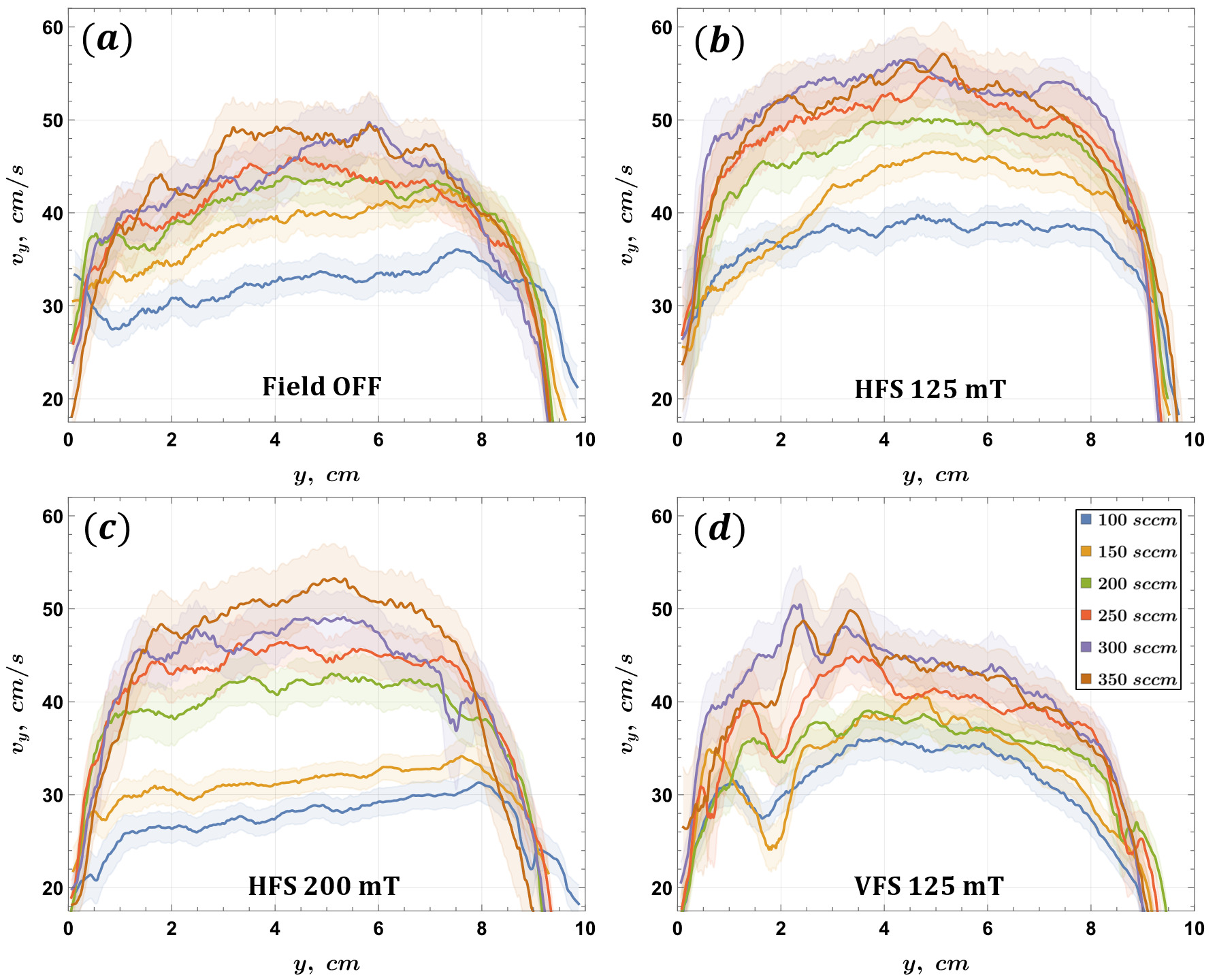}
\caption{Vertical velocity $v_y$ over elevation $y$, with mean curves and standard deviation bands.}
\label{fig:velocity-vertical-all-field-configs}
\end{center}
\end{figure}

\begin{figure}[H]
\begin{center}
\includegraphics[width=0.735\linewidth]{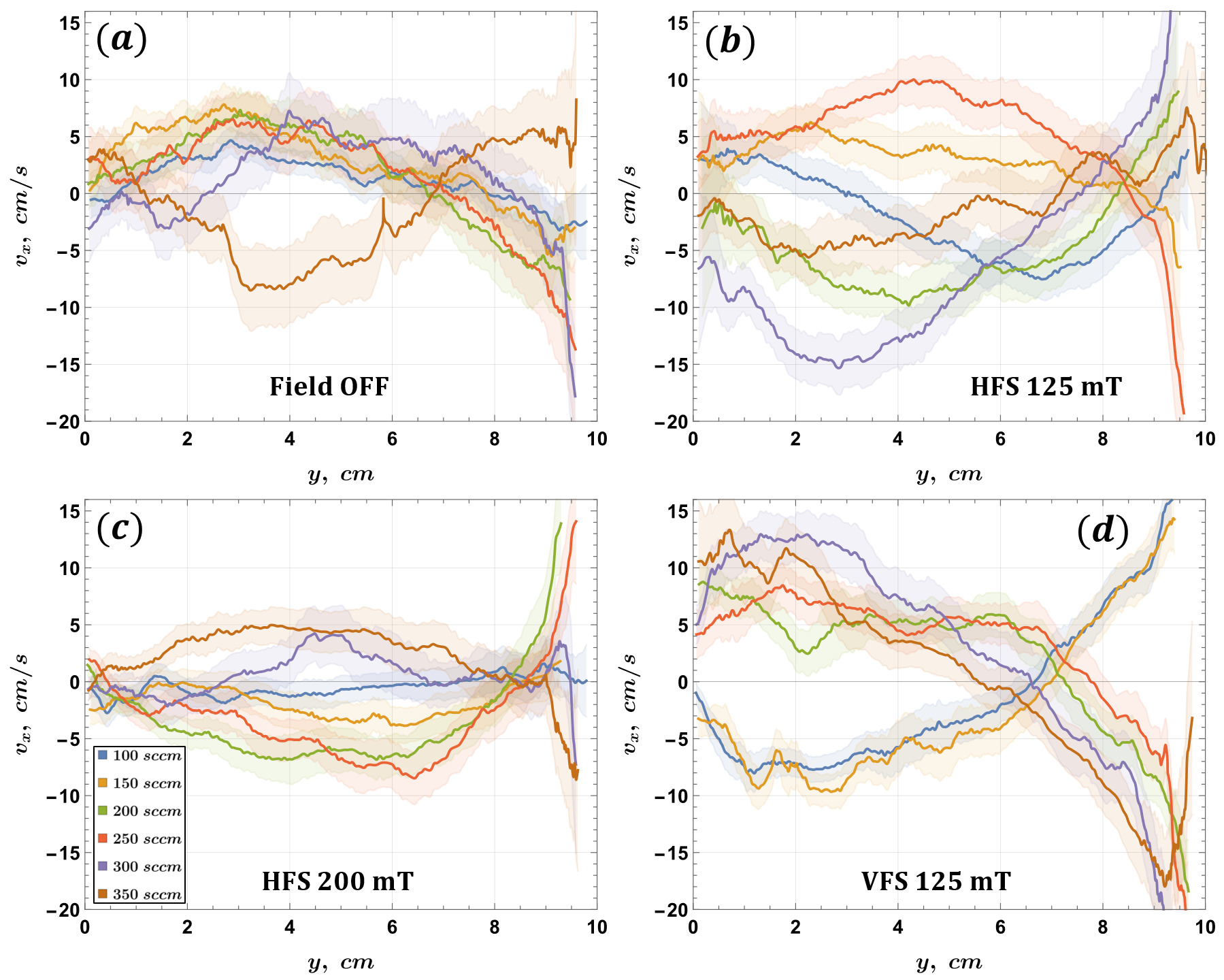}
\caption{Horizontal velocity $v_x$ over elevation $y$, with mean curves and standard deviation bands.}
\label{fig:velocity-horizontal-all-field-configs}
\end{center}
\end{figure}

Bubble velocity distribution over elevation is also of interest, and is shown in Figures \ref{fig:velocity-vertical-all-field-configs}-\ref{fig:velocity-horizontal-all-field-configs}.

\clearpage

The profiles for vertical ($v_y$) and horizontal ($v_x$) velocity are obtained from trajectories by computing displacements and centering them between the bubble coordinate points. Velocity data is filtered using QSEs \cite{antonov-qse} for outlier removal as in \cite{birjukovs2021resolving}, with identical parameters, except that mean curves are obtained by uniformly binning the $y$ range and computing bin means and standard deviations.

Figure \ref{fig:velocity-vertical-all-field-configs} shows several noteworthy trends. First, one can observe that the $v_y$ profiles with applied MF become much less oscillatory (and the standard deviation bands narrower) in the cases with HFS, and the profiles for different flow rates are more clearly differentiated, especially for $200~mT$ HFS. An interesting feature of the latter case is that, for $100$ and $150~sccm$, $v_y$ increases almost linearly in the $2$-$8~cm$ elevation range, as opposed to the other three cases where the curves clearly nonlinear. The $125~mT$ VFS case (d) is special in that $v_y$ exhibits a trend that is in opposition to the case with no MF and the cases with HFS -- $v_y$ in the $2$-$4~cm$ interval is greater than in the rest of the elevation range. It is also remarkable that bubbles on average experience a sharp velocity decrease (with a significant local minimum) after initial acceleration, then accelerate again, followed by a profile form similar to the other cases in (a-c). This suggests that, while VFS stabilizes trajectories well, it, conversely, makes the vertical bubble motion more unstable. The latter is clearly seen in comparison with (b-c), where one can see that the $v_y$ profiles are much more even compared to both the VFS case and the case without applied MF (a). Another observation is that, as the flow rate increases, the $v_y$ minimum is consistently shifted in the $y>0$ direction. Finally, note the sharp deceleration of bubbles near the free surface, which is seen for all MF configurations and across the flow rate range.

Now consider the $v_x$ profiles shown in Figure \ref{fig:velocity-horizontal-all-field-configs}. Normally, in a thicker vessel with unconfined or less confined bubble flow, one would observe very clear and consistent $v_x$ damping with decreased flow rate and increased HFS field strength within the range of MF considered here \cite{birjukovsArgonBubbleFlow2020, x-ray-validation, uttt-x-ray-single-bubble}
. Here, however, in (b) one can see that, while $v_x$ profiles are less oscillatory and more differentiated, in many cases $v_x$ is at least not smaller than with no MF. This largely stems from the bubble chain shape and motion patterns observed in Figure \ref{fig:density-bins-HFS-125mT}, and likely suggests that, due to geometric confinement and a large vortex formation to the side of the bubble chain, as well as due to turbulence damping by MF, a stable vortex/bubble chain coupled system forms that requires greater MF strength to dampen. Vortex/chain coupling for Hele-Shaw systems should be stronger than in thicker metal vessels, since momentum transferred from bubbles to the vortex is confined to a smaller metal volume. Thus, the kinetic energy density is greater, inducing a higher-velocity vortex. For $200~mT$ HFS (c), one can see that the $v_x$ maxima are much smaller (although the effect is weaker than observed for a thicker metal vessel) and not as sharp, although both (b) and (c) show horizontal acceleration near the free surface, which has already been discussed above. Another reason for this behavior is that magnetic damping of metal flow about bubbles in Hele-Shaw cells should be less effective, since the Lorentz force resulting from transverse/longitudinal field (with respect to the instantaneous bubble motion direction) is due to the closed current density loops forming about the bubbles \cite{birjukovs2021resolving, zhang-thesis, hzdr-ibm-bubbles-thesis}. These may not necessarily form properly for bubbles in contact with both walls, or even with ones in contact with a single wall. Geometric confinement can be expected to have an effect on current density loops that is somewhat similar to the principle of eddy current damping in transformer cores, where the current magnitude is reduced by confinement to thin insulated sheets. In the latter case, magnetic losses are reduced, whereas here, the effectiveness of flow kinetic energy dissipation by the Lorentz force is diminished.

In the case of $125~mT$ VFS (d), one can see that not only is $v_x$ amplified, and the maxima are shifted closer to the inlet, but there is also a reversal in $v_x$ that occurs far enough from the free surface that it cannot be explained with the same arguments as the cases with HFS. Note that these $v_x$ profiles are in line with what is seen in Figure \ref{fig:density-bins-VFS-125mT}. It is the current hypothesis of the authors that, since VFS damps the horizontal metal flow velocity component, and only affects the vertical one indirectly (flow continuity constraint), bubble chains can more readily accelerate nearby metal volume upwards, inducing a vortex. Conversely, HFS damps vertical flow, and therefore, for an equal MF strength, it promotes the vortex/chain coupling less \cite{zhang-thesis,birjukovs2021resolving}.

\section{Conclusions \& outlook}

To summarize, we have developed and demonstrated a methodology for optical imaging of bubble flow in liquid metal Hele-Shaw cells. The key feature is the side illumination system, which enables the detection of bubbles that are not translucent. The latter are still visible due to backlighting. This approach allows one to study significantly thicker Hele-Shaw systems, and should be readily applicable to up to $3$-$4.5~mm$ thick liquid metal layers, as opposed to $1~mm$ reported in the previous attempts, where one had to rely on translucent bubbles only \cite{KlaasenHeleShawMetalSolo}. The image processing code implemented to detect and track bubbles was shown to be sufficiently effective for reliable trajectory reconstruction. With the above, we have assembled a comprehensive optical imaging dataset spanning a range of flow rates, with different MF configurations (Table \ref{tab:experiment-summary}).

Based on a limited test sample from the entire recorded dataset, we have demonstrated preliminary results which enable limited physical interpretation of the observed bubble dynamics. It was shown that magnetically stabilizing bubble chain flow in a Hele-Shaw cell requires stronger MF than in thicker cells. We have quantified the effects of varying flow rate and MF on bubble flow stability by computing the widths of envelopes for the observed bubble trajectories, as well as by examining the variations of mean velocity and its standard deviation over elevation above the bottom of the vessel. An important observation is that vertical MF reduces the overall spread of bubble trajectories better (i.e., preserves bubble motion coherency more effectively with increasing flow rate) than horizontal MF of the same strength, but in turn destabilizes the motion of individual bubbles (induces strong perturbations in vertical velocity), and produces less rectilinear trajectories. Vertical MF also makes the bubble trajectory spread width vary more evenly and monotonously over the cell height. The results also suggest that, for the weaker horizontal MF and particularly for the case with vertical MF, a coupled vortex/bubble chain system forms, which determines bubble chain shapes. We hypothesize that this stems from greater density of flow kinetic energy imparted from the vertically travelling bubble chain to the surrounding metal volume, which is due to the geometric confinement of metal flow in the Hele-Shaw cell.

The authors would like to reiterate that, as intriguing and apparently physically adequate as these preliminary results are, all of the above speculations/hypotheses/observations must and will be verified again in a follow-up paper, where more bubble sequences per flow rate and MF configurations will be processed.

Regarding future plans, they are as follows. First, it is planned to complete the optical imaging dataset by performing experiments with $75~mT$ HFS/VFS systems, thus mirroring the neutron radiography image dataset acquired in \cite{birjukovs2021resolving}, but for a $3~mm$ thick metal cell, as opposed to $30~mm$. In addition, we aim to validate the imaging and data processing methods used herein against X-ray imaging experiments with $3~mm$ cells (not yet published). In addition, data is available (but in some cases not yet published) for $6$-, $12$- and $20~mm$ thick systems for a wide range of flow rates \cite{x-ray-bubble-breakup, x-ray-bubble-coalescence}. For, $3$ and $6~mm$, there are image sequences recorded with the same flow rates and MF systems as in this paper. Thus, the objective is to process the neutron, X-ray and optical imaging datasets, and study bubble chain flow behavior not only in the Reynolds/Eötvös/Stuart number space (the magnetic Reynolds number can be assumed $\ll 1$, unless flow rates are especially high), but also investigate how the observed physics depend on the degree of geometric confinement.

More specifically, both for optical imaging and the rest of the above-mentioned data, we plan to evaluate the frequency of bubble split/merge events, velocity/size statistics for the involved bubbles, as well as to observe how these events are distributed spatially. In addition to the bubble trajectory envelope thickness and velocity profiles, other metrics of interest include velocity \cite{hele-shaw-bubble-swarm-1, hele-shaw-bubble-swarm-2} and trajectory curvature probability density functions, since these quantify how and at what scales the MF and the flow rate setting affects bubble flow. The shapes of bubbles and their chains are also important to quantify, since these are coupled to metal flow about the bubbles and in the bulk metal volume. It is expected that the above will help establish potentially useful empirical relations, (in-)stability criteria, and help improve numerical models, such as the ones based on a Lagrangian approach for bubble motion and statistical treatment of bubble interactions.

The code for image acquisition, image processing as well as the 3D geometry files for the side illumination device are available on \textit{GitHub}:

\begin{itemize}[noitemsep]
    \item \href{https://github.com/ajegorovs/Grablink-Full-sequence-acquisition/tree/master}{ajegorovs/Grablink-Full-sequence-acquisition} (image acquisition)
    \item \href{https://github.com/ajegorovs/bubble_process_py}{ajegorovs/bubble\_process\_py} (image/data processing)
    \item \href{https://github.com/ajegorovs/optical_mhd_hele_shaw_cell}{ajegorovs/optical\_mhd\_hele\_shaw\_cell} (side illumination frame 3D files)
\end{itemize}

\section*{Acknowledgments}

This research is a part of the ERDF project ”Development of numerical modelling approaches to study complex multiphysical interactions in electromagnetic liquid metal technologies” (No. 1.1.1.1/18/A/108). The authors would also like to express their gratitude to Tobias Lappan and Sven Eckert (Helmholtz-Zentrum Dresden-Rossendorf, Germany) for providing acrylic vessels for the experiments, as well as to Edgars Jegorovs for assistance with the initial imaging system prototypes.

\printbibliography[title={References}]

\clearpage

\section*{Appendices}

\subsection*{\textlabel{Appendix A}{appendix:A}: Metal oxidation -- issues and troubleshooting}

\begin{figure}[htbp]
\begin{center}
\includegraphics[width=0.7\linewidth]{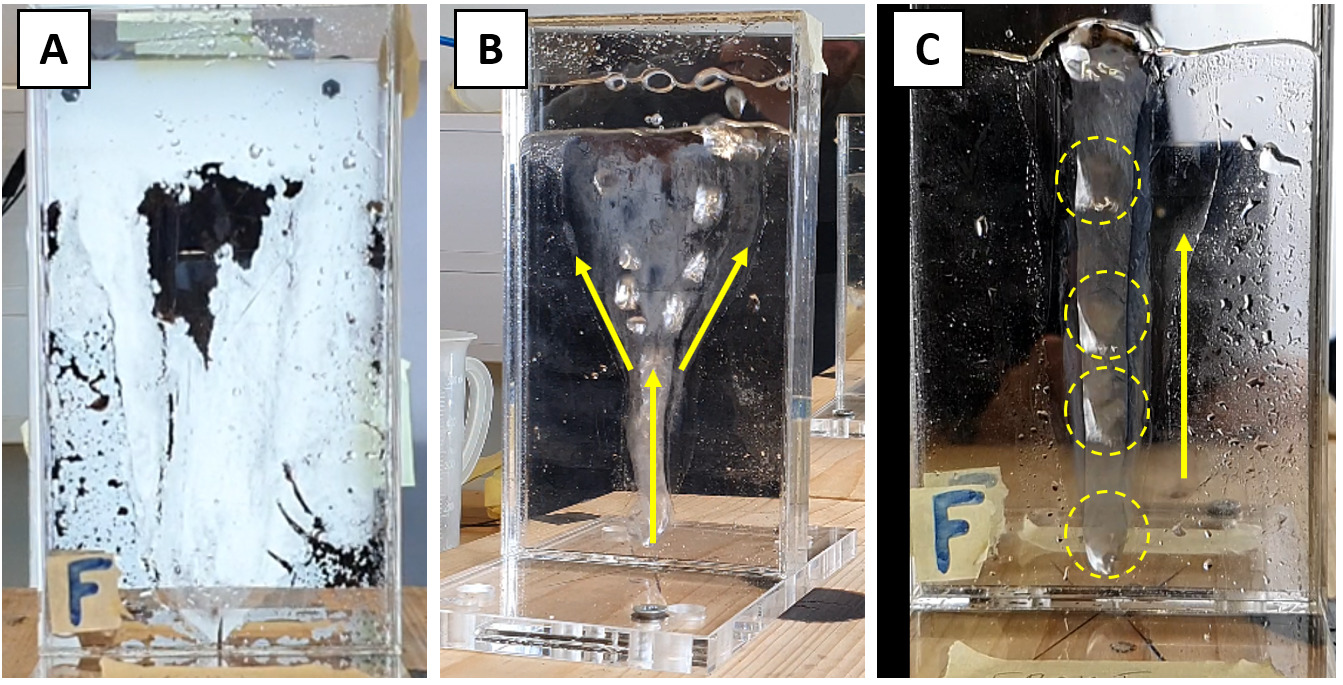}
\caption{(A) oxide deposition over the vessel walls, (B) bubble chain splitting due to interaction with an oxide film, and (C) bubble chain confinement to a path eroded through an oxide film.}
\label{fig:oxidation}
\end{center}
\end{figure}

Ga oxide deposition plays a considerable negative role in the current experiments. Oxides are dissolved in liquid metal, and then are deposited on cell walls during metal mixing. At first, a hazy film appears on internal vessel surfaces, which later develops into fully opaque thin patches, which prevent optical imaging (Figure \ref{fig:oxidation}A). We can also observe that ascending bubbles can interact with oxide patches in two ways. Oxide films likely affect surface wettability, and the effect might strong enough to force bubble splitting (Figure \ref{fig:oxidation}B). Another observed effect is the formation of long, steady state bottom-to-top bubble paths through the oxide layer, which restricts bubble motion to these paths. The flow regime also changes to low volume bubbles with abnormally small spacing in the bubble chain (Figure \ref{fig:oxidation}C).

When working with Ga or GaInSn, it is common to use HCl solution, which dissolves Ga oxide. Low concentration ($\sim 25\%$) HCl acid is readily available. The metal free surface is usually covered with a small volume of acid at stock concentration. Note that one must take care to cover the top of the liquid metal vessel, especially when imaging at higher flow rates, to avoid acid droplet ejection. Prior to using glass for front- and backplates, we performed the same experiments using fully acrylic vessels, which seemed chemically unaffected by acid. Oxide deposition is observed for both glass and acrylic surfaces. The difference is that acrylic surfaces are soft, easy to scratch during washing/rinsing, and scratches tend to accelerate the oxide deposition rate, and can result in permanent degradation of vessel translucency. There are two mechanical ways to prevent/reduce oxide deposition. First, the vessel should be emptied and rinsed with acid, which can take 2 to 10 minutes; it is possible to manually clean the vessel from within, but we advise against it, if avoidable. Second, increasing the gas flow rate during downtime (i.e., when not recording images) promotes metal stirring, which, while aerating the vessel contents, is observed to help remove the oxide.

\clearpage

\subsection*{\textlabel{Appendix B}{appendix:B}: Verifying the camera frame rate}

While the camera is coupled with the image capture card, the combination of utilized hardware has no built-in support for recording image sequences. To do this, example scripts for programmatic camera control were modified to include the required memory management operations. To ensure that modifications and camera work properly at user-specified frame rates, a rudimentary test was devised.

\begin{figure}[htbp]
\begin{center}
\includegraphics[width=0.625\textwidth]{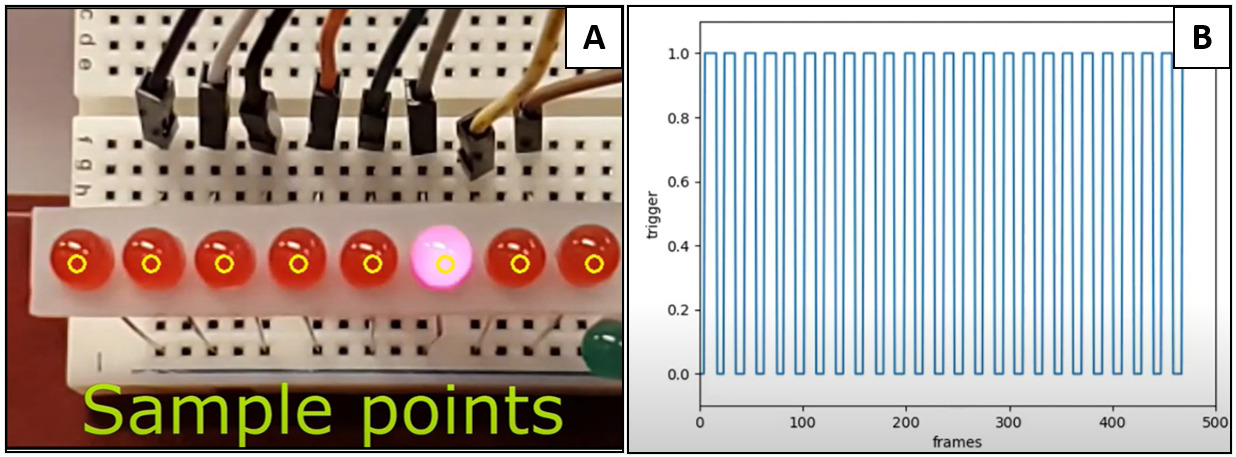}
\caption{(A) An breadboard with a row of diodes, one of which is activated, and (B) diode activation periods ($1 = \text{active}$) as imaged by the camera.}
\label{fig:exp_illum_test}
\end{center}
\end{figure}

\begin{figure}[htbp]
\begin{center}
\includegraphics[width=0.9\linewidth]{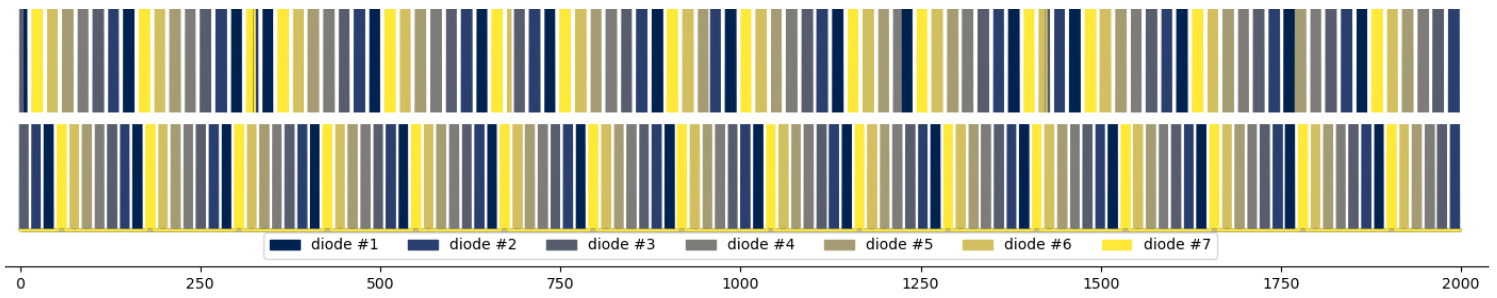}
\caption{Activation periods for different diodes (color coded by ID) and their order of activation. The top image is an example of how a camera that skips frames during recording would perform, the bottom image illustrates the correct behavior, which is what was measured for the utilized camera.}
\label{fig:exp_illum_test2}
\end{center}
\end{figure}

The test is based on recording of a row of diodes which are lit up sequentially at known time intervals. Diodes are controlled using an \textit{Arduino} board. Camera FPS is derived  by calculating the ratio of the number of frames per diode activation cycle to real time per cycle \ref{fig:exp_illum_test}. Fast Fourier transform was used to analyze the activation cycles. To determine whether the camera systematically skips frames, a separate diode activation order and activation duration graph can be analyzed \ref{fig:exp_illum_test2}. Camera behavior can be visually inspected by assigning different color codes to the detected diodes signals.

\clearpage

\subsection*{\textlabel{Appendix C}{appendix:C}: Data interpolation \& extrapolation}

\begin{figure}[htbp]
\begin{center}
\includegraphics[width=1\linewidth]{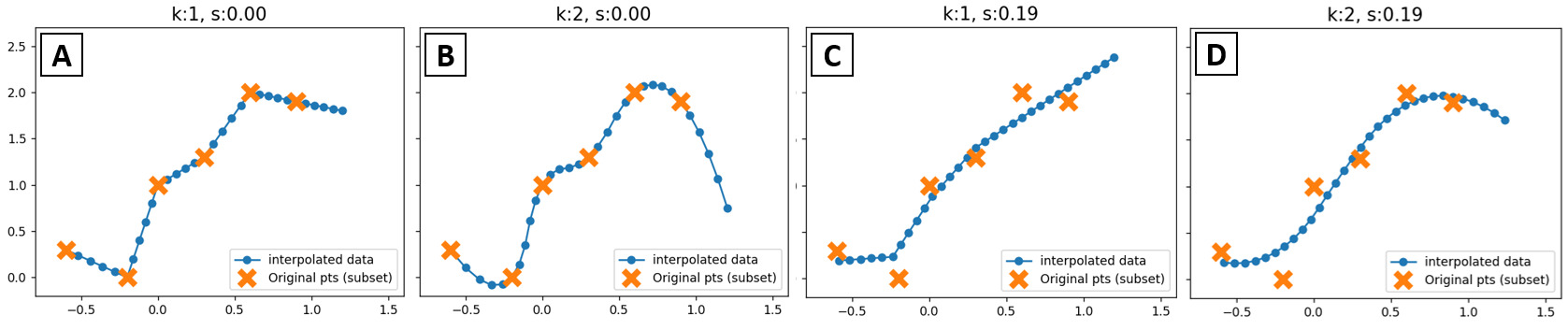}
\caption{An example of 2D trajectory extrapolation using B-splines of order $k$ and a smoothing factor $s$: (A) $k$ = 1 and $s$ = 0, (B) $k$ = 2 and $s$ = 0, (C) $k$ = 1 and $s$ = 0.19, (D) $k$ = 2 and $s$ = 0.19.}
\label{fig:extrapolation}
\end{center}
\end{figure}

For all data interpolation tasks, we use the B-spline fitting algorithm from the \textit{SciPy} library. The default \textit{BSpline} method generates splines of order $k$ for the data nodes, with enforced curve continuity and smoothness (Figure \ref{fig:extrapolation}). Polynomials are prone to excessive oscillation if data is unevenly spaced. In the current implementation, this is mitigated by using a smoothing factor $s$ which relaxes the requirement for the spline end points to coincide with data points.

Smoothing also enables the \textit{BSpline} method to be used in data extrapolation. Without smoothing, the last spline is only constrained to satisfy smoothness about previous splines, and extrapolated data rapidly diverges (Figure \ref{fig:extrapolation}B). A heavily smoothed B-spline represents a general trend in the data, but with optimized parameters yield a good extrapolation (Figure \ref{fig:extrapolation}C,D).

\begin{figure}[htbp]
\begin{center}
\includegraphics[width=0.95\linewidth]{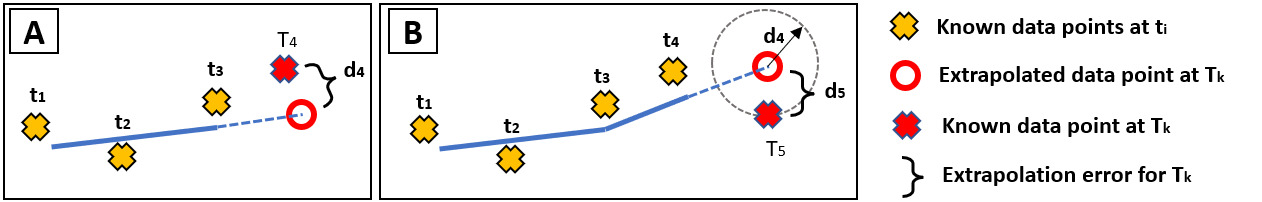}
\caption{(A) Estimation of the extrapolation error $d_4$, which can be used to evaluate extrapolation quality in (B). $T_k/t_k$ denote time stamps.}
\label{fig:extrapolation2}
\end{center}
\end{figure}

To choose appropriate B-spline parameters, their different combinations are tested on known data (resolved parts of trajectories) and the best performing parameter pair is kept. To quantitatively describe extrapolation performance, we examine subsets of a trajectory, drop the endpoint, and try extrapolating its position (Figure \ref{fig:extrapolation2}A,B). The difference $d$ between extrapolated and actual positions is the extrapolation error, which is minimal for the best $(s,k)$ values. Additionally, these tests are used to estimate thresholds used in trajectory reconstruction.

\end{document}